\documentclass[pra,twocolumn,showpacs,amsmath,amssymb,superscriptaddress]{revtex4-2}
\usepackage{braket}
\usepackage{cases}
\usepackage{color}
\usepackage{graphicx,subfigure}
\usepackage{float}
\usepackage{tikz}
\usepackage{amsmath}
\usetikzlibrary{quantikz}
\usepackage{bm}
\usepackage{hyperref}
\hypersetup{
    colorlinks=true, 
    linkcolor=cyan,
    citecolor=magenta, 
    filecolor=magenta, 
    urlcolor=cyan,
    runcolor=cyan
}

\newcommand{\ignore}[1]{}

\usepackage{commath}
\usepackage{color}
\usepackage{amsmath}
\usepackage{adjustbox}
\usepackage[normalem]{ulem}

\def\be{\begin{equation}}
\def\ee{\end{equation}}

\newcommand{\bes} {\begin{subequations}}
\newcommand{\ees} {\end{subequations}}
\newcommand{\beq}{\begin{equation}}
\newcommand{\eeq}{\end{equation}}

\def\a{\alpha}
\def\b{\beta}
\def\g{\gamma}
\def\d{\delta}

\def\r{\rho}     
\def\s{\sigma}

\def\o{\omega}

\def\ox{\otimes}
\def\>{\rangle}
\def\<{\langle}
\def\Tr{\mathrm{Tr}}

\newcommand{\ketb}[2]{|{#1}\>\!\<#2|}

\begin{document}

\title{Suppression of crosstalk in superconducting qubits using dynamical decoupling}

\author{Vinay Tripathi}
\affiliation{Department of Physics \& Astronomy, University of Southern California,
Los Angeles, California 90089, USA}
\affiliation{Center for Quantum Information Science \& Technology, University of
Southern California, Los Angeles, California 90089, USA}
\author{Huo Chen}
\affiliation{Center for Quantum Information Science \& Technology, University of
Southern California, Los Angeles, California 90089, USA}
\affiliation{Department of Electrical \& Computer Engineering, University of Southern California,
Los Angeles, CA 90089, USA}
\author{Mostafa Khezri}
\affiliation{Center for Quantum Information Science \& Technology, University of
Southern California, Los Angeles, California 90089, USA}
\affiliation{current address: Google Quantum AI}
\author{Ka-Wa Yip}
\affiliation{Department of Physics \& Astronomy, University of Southern California,
Los Angeles, California 90089, USA}
\affiliation{Center for Quantum Information Science \& Technology, University of
Southern California, Los Angeles, California 90089, USA}
\author{E.~M.~Levenson-Falk}
\affiliation{Department of Physics \& Astronomy, University of Southern California,
Los Angeles, California 90089, USA}
\affiliation{Center for Quantum Information Science \& Technology, University of
Southern California, Los Angeles, California 90089, USA}
\author{Daniel A. Lidar}
\affiliation{Department of Physics \& Astronomy, University of Southern California,
Los Angeles, California 90089, USA}
\affiliation{Center for Quantum Information Science \& Technology, University of
Southern California, Los Angeles, California 90089, USA}
\affiliation{Department of Electrical \& Computer Engineering, University of Southern California,
Los Angeles, CA 90089, USA}
\affiliation{Department of Chemistry, University of Southern California, Los Angeles,
CA 90089, USA}

\date{\today}
\begin{abstract}
Currently available superconducting quantum processors with interconnected transmon qubits are noisy and prone to various errors. The errors can be attributed to sources such as open quantum system effects and spurious inter-qubit couplings (crosstalk). The $ZZ$-coupling between qubits in fixed frequency transmon architectures is always present and contributes to both coherent and incoherent crosstalk errors. Its suppression is therefore a key step towards enhancing the fidelity of quantum computation using transmons.
Here we propose the use of dynamical decoupling to suppress the crosstalk, and demonstrate the success of this scheme through experiments performed on several IBM quantum cloud processors. In particular, we demonstrate improvements in quantum memory as well as the performance of single-qubit and two-qubit gate operations.
We perform open quantum system simulations of the multi-qubit processors and find good agreement with the experimental results. We analyze the performance of the protocol based on a simple analytical model and elucidate the importance of the qubit drive frequency in interpreting the results. In particular, we demonstrate that the XY4 dynamical decoupling sequence loses its universality if the drive frequency is not much larger than the system-bath coupling strength. 
Our work demonstrates that dynamical decoupling is an effective and practical way to suppress crosstalk and open system effects, thus paving the way towards higher-fidelity logic gates in transmon-based quantum computers.

\end{abstract}

\maketitle

\textit{Introduction}.---%
Quantum computing is currently at a stage where noisy gate-model devices with only a few dozens of qubits have enabled an exploration of various algorithms and quantum information protocols~\cite{Preskill:2018aa}. Among the leading implementations in this field are superconducting transmon qubits~\cite{Koc07a,Clarke2008}, designed to have a suppressed sensitivity to charge noise and thus possessing a higher coherence and lifetime than most other superconducting qubit types~\cite{Kjaergaard_2020}. 
 Transmons have been used to demonstrate quantum information processing in a series of recent experiments~\cite{Hacohen2016,Kandala:2017aa,Minev2019,Arute:2019aa,Havlicek2019,Kandala2019,Campagne2020,Andersen2020,Arute2020,wu2021strong,ying2021floquet}. 

In order to construct a large scale quantum computer (QC) capable of performing useful tasks it must be possible to both store and process quantum information with sufficiently low error rates so as to perform fault-tolerant quantum computation~\cite{Aliferis:05,Chao:2017ab,Campbell:2017aa}. These requirements are affected by different types of errors which afflict quantum processors. Primary error sources are open quantum system effects resulting in decoherence, and spurious coupling between qubits, control lines, and readout apparatus, resulting in crosstalk. For fixed frequency transmon processors, the $ZZ$-coupling between any two neighboring qubits is always present and contributes to both coherent and incoherent errors, making the suppression of $ZZ$-crosstalk one of the most important challenges for such processors. Various schemes have been proposed and demonstrated toward this end, e.g., 
combining a capacitively shunted flux qubit and a transmon qubit with opposite-sign anharmonicity~\cite{Ku2020,Zhao2020}. For qubits coupled via tunable couplers, it is possible to adjust the coupler frequency such that the $ZZ$ interactions from each coupler destructively interfere~\cite{Mundada2019}, or to detune neighboring qubits~\cite{chu2021couplerassisted}. However, a simple and universal scheme for $ZZ$-crosstalk suppression based purely on transmons is still lacking. Here we propose and experimentally demonstrate such a scheme using three different 5-qubit IBM Quantum Experience (IBMQE) processors~\cite{IBMQE}. Our scheme is based on dynamical decoupling (DD)~\cite{Viola:98,Vitali:99,Duan:98e,Zanardi:1999fk} -- the simplest of all quantum error correction or suppression protocols~\cite{Lidar-Brun:book}. 

We demonstrate that DD is highly effective at suppressing $ZZ$-crosstalk, while at the same time also suppressing unwanted system-bath interactions responsible for decoherence. Previous work on the use of DD to protect transmon qubit states did not separate these two different contributions to fidelity decay~\cite{Pokharel2018,souza2020process,jurcevic2020demonstration}, and we show here that fidelity oscillations that were previously interpreted as a possible sign of a non-Markovian bath are in fact attributable to crosstalk.
We demonstrate these results experimentally and numerically, and provide an analytical basis for the choice of the pertinent DD pulse sequences. We highlight the important role played by gate calibration procedures in interpreting the outcomes of our crosstalk suppression experiments. We then use our DD scheme to improve the performance of both single-qubit and two-qubit gate operations by suppressing the crosstalk originating from the spectator qubits.

\begin{figure*}[t]
\subfigure{\includegraphics[width=.95\linewidth]{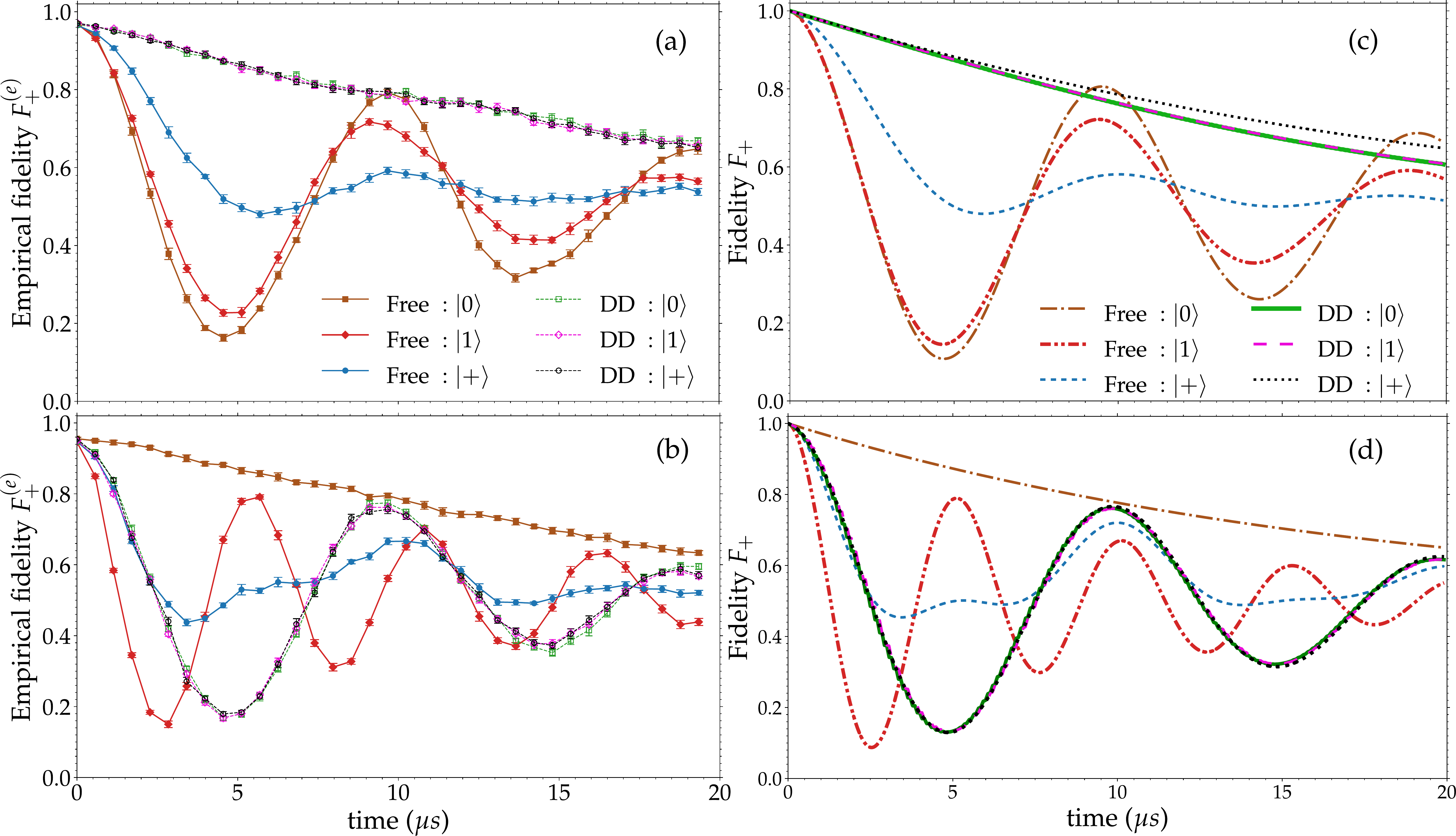}}
\caption{Fidelity of the $\ket{+}$ state of the main qubit, for different spectator qubits' initial states $\{\ket{0},\ket{1},\ket{+}\}$, obtained experimentally [(a) and (b)] and by solving the Redfield master equation [(c) and (d)]. (a) Results for Ourense. The free evolution curves all exhibit oscillations with nearly equal periods but distinct amplitudes. These effects disappear under the application of the XY4 sequence just to the spectator qubits, leaving only a common fidelity decay. 
(b) Results for Yorktown. The free evolution curves range from monotonic to oscillatory. The differences disappear under DD applied to the spectator qubits, leaving only a common damped oscillation.  
Error bars denote $95\%$ confidence intervals. (c) Redfield equation simulation results for a multi-qubit system with $\omega_{\rm d}=\omega_{q_1}$.
(d) Simulation results for $\omega_{\rm d}=\omega_{q_1}-2J$.
All qualitative features observed in (a) and (b) are reproduced in (c) and (d), respectively. The crosstalk strength considering a two-qubit model is (c) $J/2\pi=51.55$\;KHz and (d) $J/2\pi=52.63$\;KHz,
obtained by fitting the periods of (a) and (b).
}
\label{fig:main_1}
\end{figure*}

\textit{Experimental results for state protection}.---%
We consider Ramsey-like experiments, where we prepare an initial state, let it evolve, and then undo the state preparation. Our experiments were conducted on the \emph{ibmq\_ourense} (Ourense), \emph{ibmq\_5\_yorktown} (Yorktown), and \emph{ibmq\_lima} (Lima) five-qubit IBMQE processors. In each case we selected one ``main'' qubit and consider the other four to be ``spectator" qubits. We performed two types of experiments: \emph{free} and \emph{DD-protected} evolution. 

In both cases the main qubit was initialized in the $\ket{0}$ state, then an $R_y(\pi/2)$ gate was applied to prepare it in the $\ket{+}=(\ket{0}+\ket{1})/\sqrt{2}$ state (other initial states are discussed in Ref.~\cite[Sec.~B]{SM}).
All spectator qubits were initialized in the same state, which we varied. In the free evolution case we then applied a series of identity gates (separated by barriers) on all the qubits, followed by $R_y(-\pi/2)$ on the main qubit to undo the $\ket{+}$ state preparation. In the DD-protected case we applied the universal XY4 sequence~\cite{Maudsley:1986ty,Viola:99,Souza:2012aa} to all the spectator qubits, but only identity gates to the main qubit. The ideal XY4 sequence comprises repetitions of $X f_\tau Y f_\tau X f_\tau Y f_\tau$, where $f_\tau$ denotes free evolution for a duration of $\tau$ (in our experiments $\tau=71.1$\;ns), and $X$ and $Y$ are instantaneous $\pi$ pulses about the $x$ and $y$ axes, respectively (in reality, the pulses have a finite duration; we spaced them without any delay, so $\tau$ is their peak-to-peak spacing). This was done primarily in order to decouple the $ZZ$-crosstalk term; it also  suppresses undesired system-bath interactions, as explained below. After each gate sequence, we measured the main qubit in the computational ($0/1$) basis. Each circuit was repeated $8192$ times and we used the fraction of $0$ outcomes on the main qubit, $F_+^{(e)}$, as the empirical fidelity measure (augmented by bootstrapping; see Ref.~\cite[Sec.~A]{SM}), i.e., as a proxy for $F_+\equiv\Tr\{R_y(-\pi/2)\mathcal{E}[R_y(\pi/2)\ketb{0}{0}R_y(-\pi/2)]R_y(\pi/2)\}$, where $\mathcal{E}$ is the quantum map corresponding to either free or DD-protected evolution of the main qubit. Ideally $R_y(\pi/2)\ket{0}=\ket{+}$ and $\mathcal{E}$ corresponds to the identity channel; in reality $R_y(\pi/2)$ prepares a slightly different state due to gate errors, and $\mathcal{E}$ corresponds to a noisy channel.

Fig.~\ref{fig:main_1} shows the results of the free evolution and DD-protected evolution experiments on Ourense (panel a) and Yorktown (panel b). We plot the empirical fidelity $F_+^{(e)}(t)$ of the $\ket{+}$ state of the main qubit for different spectator qubits' initial states $\{\ket{j}\}_{j=0,1,+}$. The envelope of the DD-protected evolution decays more slowly than that of the free evolution, for both processors (see also Ref.~\cite[Sec.~B]{SM}). A glance at the Ourense and Yorktown results reveals striking differences for two identical sets of experiments. We explain below how these arise due to the different choice of qubit drive frequency for the two processors. Conversely, a striking qualitative similarity between the two is that the DD-protected evolution essentially erases the difference between the three different spectator initial states, whereas the differences are pronounced in the free evolution case.

Having confirmed that the DD protocol works as intended, we next provide a theoretical explanation for our findings. 
Additional confirmation of the efficacy and robustness of DD in suppressing crosstalk is provided in Ref.~\cite[Sec.~B]{SM}.

\textit{Model}.---%
Consider for simplicity a system of just two nominally uncoupled qubits which, however, have an undesired and always on $ZZ$-coupling of strength $J\neq 0$, just like coupled transmons. The generalization to $n>2$ qubits is straightforward and is considered in Ref.~\cite[Sec.~C]{SM}. The effective Hamiltonian can be written as
$H_{\rm S}=-\frac{\omega_{q_1}}{2} Z_1-\frac{\omega_{q_2}}{2} Z_2+J ZZ$,
where $\omega_{q_1}$ and $\omega_{q_2}$ are the qubit frequencies of the main and the spectator qubit respectively, 
where $Z_2\equiv IZ\equiv I\otimes\s^z$, etc.
The $ZZ$-coupling term dresses the qubit frequencies such that the frequency of the main qubit changes and depends on the state of the spectator qubit. Correspondingly, we define the eigenfrequencies of the main qubit by 
the spectral gap of $H_{\rm S}$ after
fixing the state of the spectator qubit to either $\ket{0}$, $\ket{1}$, or $\ket{+}$ (i.e., replacing the spectator qubit operators in $H_{\rm S}$ by their expectation values $\{-1,1,0\}$). 
This yields, respectively, $\o_{\rm eig}^0 
= \omega_{q_1}-2J$, $\o_{\rm eig}^1 
= \omega_{q_1}+2J$, and 
$\o_{\rm eig}^+ 
=\omega_{q_1}$, which is also the bare frequency of the main qubit. In Ref.~\cite[Sec.~D]{SM} we show that this conclusion is identical to one derived from a first-principles model of transmons as multi-level systems.
We show below that these different eigenfrequencies explain the difference between the Ourense and Yorktown processors seen in Fig.~\ref{fig:main_1}.

In the open system settings, the total Hamiltonian can be written as 
$H=H_{\rm S} + H_{\rm SB}$,
where we 
assume a general system-bath interaction
$H_{\rm SB} = \sum_{\alpha,\beta\in \{0,x,y,z\}}g_{\alpha \beta} \s^{\alpha}\otimes\s^\b\otimes B_{\alpha\beta}$,
where $\s^{0}=I$, 
$g_{\alpha\beta}=g_{\beta\alpha}^*$ is
the strength of the coupling to the bath, and $B_{\alpha\beta}$ are Hermitian bath operators. 

Let us now move to a rotating frame defined by the number operator $\hat{N}
= I-\frac{1}{2}(Z_1+Z_2)$ for the main and spectator qubits. Ignoring the overall energy shift, we write the unitary transformation operator $U(t) = e^{i\omega_{\rm d} \hat{N}t}$, where $\omega_{\rm d}\neq 0$ is the drive frequency, as 
$U(t) = e^{-i\omega_{\rm d}(Z_1+Z_2)t/2}$.
The rotating frame Hamiltonian $\tilde{H}(t) = U H U^\dag + i \dot{U} U^\dag$ becomes
\begin{align}
\label{eq:Heff}
    \tilde{H}(t) = \sum_{i=1}^2 \Omega_i Z_i + J ZZ + \tilde{H}_{\rm SB}(t)\ ,\ \Omega_i \equiv \frac{\omega_{\rm d}-\omega_{q_i}}{2}
\end{align}
where $\tilde{H}_{\rm SB}(t) = \sum_{\alpha\beta}g_{\alpha \beta} \left[U(t)\left(\s^{\alpha}\otimes\s^\b\right)U^{\dagger}(t)\right]\otimes B_{\alpha\beta}$.
The free evolution unitary generated by the rotating frame Hamiltonian is 
$\tilde{U}_f(t_f,t_i) = \mathcal{T}_{+}\exp\left(-i\int_{t_i}^{t_f}dt\ \tilde{H}(t) \right)$, 
with $\mathcal{T}_{+}$ denoting forward time-ordering.

\textit{Free evolution and extraction of the crosstalk frequency}.---%
The choice of $\o_{\rm d}$ gives rise to different rotating frame Hamiltonians.
Consider the rotating frames corresponding to the different eigenfrequencies mentioned above: $\omega_{\rm d} \in \{\o_{\rm eig}^0,\o_{\rm eig}^1,\o_{\rm eig}^+\}$.
Let $\Delta = \omega_{q_1}-\omega_{q_2}$ denote the detuning between the two qubit frequencies, and $s\in\{0,1,+\}$ ($s$ for spectator)
the rotating frame according to the choice of $\omega_{\rm d}$. 
Up to a constant, the system-only Hamiltonians obtained from $\tilde{H}(t)$ [Eq.~\eqref{eq:Heff}] in the two frames that are experimentally realized in the IBMQE devices we used are 
\bes
\label{eq-H_eig}
\begin{align}
\tilde{H}_{\rm S}^{+} &= - (\Delta+2J)\ketb{01}{01} -2J\ketb{10}{10} -\Delta \ketb{11}{11} \label{eq-H_eig1} \\
\tilde{H}_{\rm S}^{0} &= - \Delta \ketb{01}{01} + (4J-\Delta) \ketb{11}{11}
\label{eq-H_eig2} \ .
\end{align}
\ees
Let us furthermore assume a simple Markovian dephasing model, with a Lindbladian of the form $\mathcal{L}^s = -i[\tilde{H}_{\rm S}^s,\cdot]+\sum_\a \g_\a (L_\a\cdot L_\a^\dag -\frac{1}{2}\{L_\a^\dag L_\a,\cdot\})$, where the Lindblad operators are $\{L_1 = ZI,L_2 = IZ,L_3 = ZZ\}$. Let $\g = \g_1+\g_3$.
Using Eq.~\eqref{eq-H_eig}, 
under free evolution the probability $p^{s}_{+s'}(t)$ of the main qubit's final state being $\ket{+}$ if its initial state is $\ket{+}$ (i.e., $F_+$), is, for the three different initial state $\ket{s'}$ $s'\in\{+,0,1\}$ of the spectator qubit (see Ref.~\cite[Sec.~E]{SM}):
\bes
\label{eq:p}
\begin{align}
p^{+}_{+s'}(t) &= \frac{1}{2}\left(1+e^{-2\g t}\cos(2Jt)\right) \quad \forall s'\in\{+,0,1\}
\label{eq:p-bare}\\
p^{0}_{+s'}(t) &= \frac{1}{2}\left(1+e^{-2\g t}f_{s'}(t)\right)\\
f_+(t) &= \cos^2(2Jt)\ , \ f_0(t)=1\ , \ f_1(t) = \cos(4Jt)
\label{eq:p-eig0_1}
\end{align}
\ees
In the $s=+$ frame we thus expect to observe damped fidelity oscillations with a period of $\tau^+ =2\pi/2J$ for all spectator states, consistent with our data in Fig.~\ref{fig:main_1}(a).
Likewise, in the $s=0$ frame we expect no oscillations in the $\ket{0}$ case, but damped fidelity oscillations with a period of $\tau^{0}= 2\pi/4J$ when the spectator qubit is prepared in the $\ket{1}$ or $\ket{+}$ states, consistent with Fig.~\ref{fig:main_1}(b). The larger amplitude oscillations observed in Fig.~\ref{fig:main_1}(b) for the $\ket{1}$ case are also in agreement with Eq.~\eqref{eq:p-eig0_1}. We thus conclude that the Ourense and Yorktown main qubit drive frequencies are $\omega_{q_1}$ and $\o_{q_1}-2J$, respectively, i.e., the devices were calibrated with the spectator qubits in $\ket{+}$ and $\ket{0}$, respectively. Moreover, the oscillations predicted by Eq.~\eqref{eq:p} are entirely crosstalk-induced (they disappear when $J=0$).
Note that while Eq.~\eqref{eq:p-bare} incorrectly predicts equal amplitude oscillations for all three initial spectator states, this is remedied by including a lowering operator $\ketb{0}{1}$ as an additional Lindblad operator; however, this model fails to capture the observed ordering of the fidelity amplitudes (see Ref.~\cite[Sec.~E]{SM}).

\textit{Suppression of $ZZ$-crosstalk and system-bath interactions using dynamical decoupling}.---%
Having established that crosstalk (and not environmentally induced non-Markovian dynamics) suffices to explain the fidelity oscillations observed in our free evolution experiments, we now analyze its suppression using DD. 
For simplicity, consider a DD sequence consisting purely of ideal (i.e., zero-width) $X$-pulses of the form $i I\otimes e^{-i\frac{\pi}{2}X} = X_2$ 
(henceforth we ignore global phases) applied just to the spectator qubit.
In the rotating frame, 
the time evolution at the end of one cycle of such a ``pure-X'' DD sequence with pulse interval $\tau$ is given by
$\tilde{U}_X (2\tau) = X_{2} \tilde{U}_f(2\tau,\tau) X_{2} \tilde{U}_f(\tau,0)$.
Using $U e^{A} U^{\dagger}=e^{U A U^{\dagger}}$ ($A$ arbitrary, $U^\dag U=I$), 
we may write $X_{2} \tilde{U}_f(2\tau,\tau) X_{2} = \mathcal{T}_{+}\exp\left(-i\int_{\tau}^{2\tau}dt\ X_2\;\tilde{H}(t)X_2 \right)$.
Using the Magnus expansion (see, e.g., Refs.~\cite{Viola:99,Ng:2011dn}) one can show that, to first order in $\tau$, this sequence cancels every term in $\tilde{H}(t)$ that anticommutes with $X_2$. 
Using Eq.~\eqref{eq:Heff}, we are thus left with
$\tilde{U}_X(2\tau) =\tilde{U}' (2\tau) + O(\tau^2)$,
where 
$\tilde{U}' (2\tau) = \exp\big[-i\tau(\omega_{\rm d}-\omega_{q_1})Z_1-i\int_0^{\tau} dt [\tilde{H}_{\rm SB}(t) + X_2\tilde{H}_{\rm SB}(t+\tau)X_2]\big]$.
In the XY4 case, the integral also contains $Y_2\tilde{H}_{\rm SB}(t+2\tau)Y_2 + Z_2\tilde{H}_{\rm SB}(t+3\tau)Z_2$. 
As shown in Ref.~\cite[Sec.~F]{SM}, in both the pure-X and XY4 cases the integral always vanishes for the terms $Z_2$ and $ZZ$, as these terms anticommute with $X_2$ both in the lab frame and the rotating frame. I.e., it follows from the form of $\tilde{U}' (2\tau)$ that the $ZZ$-crosstalk 
and the $Z_2$ and $ZZ$ bath-coupling terms are all suppressed 
to $O(\tau)$ by both the ideal pure-X and XY4 sequences.
\emph{This explains the suppression of crosstalk observed in our experiments}.
The fact that in the $s=+$ frame $\omega_{\rm d}=\omega_{q_1}=\o_{\rm eig}^{+}$ means that the $Z_1$ term in $\tilde{U}' (2\tau)$ vanishes. \emph{This explains the absence of oscillations in the Ourense DD results} [Fig.~\ref{fig:main_1}(a)]. Likewise, the $Z_1$ term in $\tilde{U}' (2\tau)$ remains in the $s=0$ frame, when $\o_{\rm d} = \o_{\rm eig}^0$. \emph{This explains the remaining oscillations in the Yorktown DD results} [Fig.~\ref{fig:main_1}(b)].
Moreover, the suppression of the $Z_2$ and $ZZ$ bath-coupling terms explains why the fidelity under DD is generally higher than for free evolution,
as can be seen in Fig.~\ref{fig:main_1} (see also Ref.~\cite[Sec.~B]{SM}).

Now note that a term in $H$ that anticommutes with $X_2$ (e.g., $YY$) may transform to a term in $\tilde{H}(t)$ that does not; this causes many terms to not cancel to first order in $\tau$ under pure-X DD. As a consequence \emph{the XY4 sequence loses its exact universality in the rotating frame}. This may adversely affect the performance of DD sequences which are designed for high-order cancellation~\cite{Khodjasteh:2005xu,Uhrig:2007qf}.
Since the pure-X sequence is shorter by a factor of $2$, it is preferred in the present setting. We explain the reasons why the pure-X and XY4 sequences exhibit this non-standard behavior in Ref.~\cite[Secs.~F-H]{SM}, and discuss the conditions under which the universality of XY4 can be approximately recovered. 

\textit{Numerical results}.---%
The theoretical analysis above was oversimplified since it missed features such as the unequal decay rates associated with different initial spectator states [Eq.~\eqref{eq:p} predicts the same decay rate $\g$ for all three such states], and the fidelity amplitude ordering. 
Thus we now complement this analysis with a numerical study. We
assume a system-bath interaction with linear coupling terms: 
$H_{\rm SB} = \sum_{i=0}^{n-1} \sum_{\a\in\{x,y,z\}} g_{\a} \s_i^\a \otimes B_{i\a}$,
with all $n$ qubits independently coupled to Ohmic baths. We simulated the transmon system (modeled as qubits) via the Redfield master equation using the HOQST package~\cite{chen2020hoqst}. 
Our open system model is described in detail in Ref.~\cite[Sec.~C]{SM}. 
\begin{figure*}[t]
\includegraphics[width=.95\linewidth]{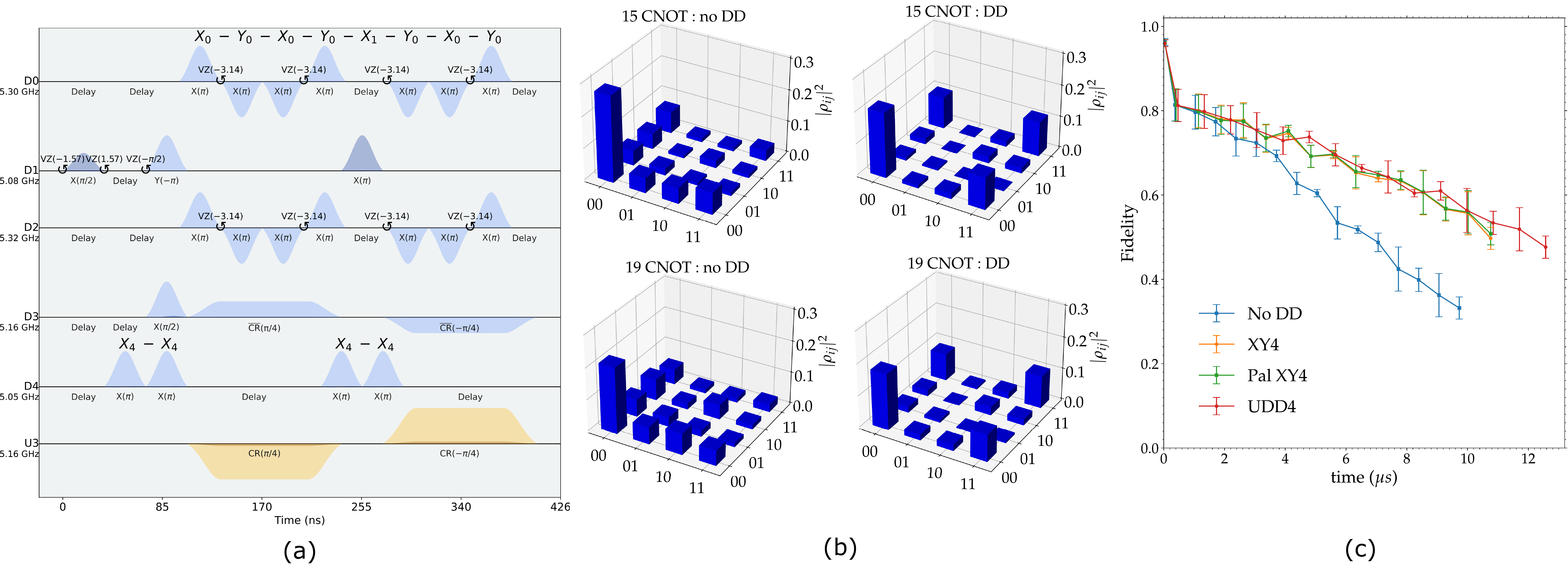}
\caption{(a) Optimized pulse and DD sequence placement in the CR-based CNOT gate. D0-D4 denote drive channels for qubits Q0-Q4. U3 represents CR pulses acting on Q1 at the Q3 frequency. The control and target qubits are Q1 and Q3, respectively; the rest are the spectator qubits. We apply the XY4 (or palindromic XY4 or UDD4) sequence to Q0 and Q2, and the pure-X sequence to Q4, with pulses placed in gaps between the CR pulses. Note that one $X$ gate in the DD sequence applied to Q0 and Q2 has been replaced by the pre-existing $X$ gate on Q1. VZ denotes the virtual $Z$ gate. (b) QST results after applying 15 (top) and 19 (bottom) CNOT gates. Left: without DD. Right: with XY4. Clearly, the XY4 results are significantly closer to the expected Bell state, i.e., equal corner peaks of $0.25$. (c) Fidelity of Bell state preparation after a repeated odd number of up to 29 CNOT gates, averaged over 5 separate runs with the spectator qubits initialized in $\ket{0}$ (we checked and found the effect of different initial spectator states to be insignificant). Error bars represent $95\%$ confidence intervals. CNOT fidelity$<1$ at $0$ gates is due to preparation and measurement errors. CNOT with DD takes longer than without DD since to avoid overlap we inserted delays to accommodate the two pure-X sequences on D4. The fidelity with DD is statistically significantly higher than fidelity without DD for all DD sequences we tried after $\sim3\mu$s, or $\sim 9$ consecutive CNOT gates.}  
\label{fig:main_2}
\end{figure*}

Figure~\ref{fig:main_1} reveals that overall, our simulations are in close qualitative agreement with the experimental results. The $\ket{+}$ curve has the smallest amplitude in our experiments, which our simulations account for by setting $g_{z} > g_{x},g_{y}$ (see Ref.~\cite[Sec.~C]{SM}). I.e., \emph{we conclude from our simulations that dephasing dominates over the other noise channels}. This is consistent with the data documented in Ref.~\cite[Sec.~A]{SM}, which shows that almost all spectator qubits have $T_2< T_1$. 
Our simulations qualitatively reproduce the oscillation pattern of the free evolution of the $\ket{+}$ initial state in Fig.~\ref{fig:main_1}(b). This required accounting for all of the spectator qubits coupled to the main qubit (unlike in the phenomenological Lindblad model with only one spectator). Thus, a multi-qubit description of the system is needed to fully understand and characterize the crosstalk in these devices. For more details see 
Ref.~\cite[Sec.~C]{SM}.

\textit{DD for gate operations}.---%
Having established the efficacy of DD in state preservation, we finally apply DD to counter crosstalk and decoherence-induced errors during gate operations; we call the resulting gates ``DD-protected gates'' (DDPGs)~\cite{Ng:2011dn}. We focus on the CNOT gate (based on cross resonance (CR) \cite{Rigetti2010,Tripathi:2019vb}) and present additional free \textit{vs} single-qubit DDPG results in Ref.~\cite[Sec.~B]{SM}, which demonstrate a significant improvement, namely, a reduction of the fidelity decay rate by more than a factor of $2$. Here we use \emph{ibmq\_quito} (Quito), a five qubit IBMQE processor. 
We choose two qubits as control and target, which we prepare in the $\ket{+,0}$ state, and the remaining three are the spectator qubits. 

Crosstalk suppression in CNOT gates has previously been explored by dividing the CNOT gate into 4 CR pulses (along with single qubit gates) and applying an $X$ pulse on all spectators after the second CR pulse~\cite{Takita2016,Takita2017}. Here, we optimize error suppression in CNOT gates by exploring a wide range of pulse placements and DD sequences. We find that the DDPG solution depicted in Fig.~\ref{fig:main_2}(a) significantly improves performance, as evidenced in Fig.~\ref{fig:main_2}(b),(c). More specifically, we apply an odd number of CNOT gates, then perform quantum state tomography (QST) and compute the fidelity with respect to the expected Bell state $(\ket{00} + \ket{11})/\sqrt{2}$. We then compare to gates integrating different types of DD sequences including XY4 (XYXY), palindromic XY4 (XYXYYXYX) and the 4th order Uhrig DD sequence (UDD4) \cite{Uhrig:2007qf} on the spectator qubits. Figures~\ref{fig:main_2}(b) and (c), respectively, show QST results and the fidelity with and without DD. Clearly, incorporation of XY4 yields a notable improvement in CNOT performance, as a function of the number of consecutive CNOT gates. The results for palindromic XY4 and UDD4 are statistically indistinguishable from XY4  (see Ref.~\cite[Sec.~B]{SM}).

\textit{Conclusions}.---%
We have formulated and experimentally implemented a simple, effective DD scheme to suppress $ZZ$-crosstalk in a multi-qubit transmon processor, that unlike other approaches~\cite{Ku2020,Zhao2020,Mundada2019} does not require any hardware redesign. 
The same scheme also suppresses interactions with the ambient bath, resulting in a significant improvement in quantum memory and gate performance. By removing crosstalk, it becomes possible to achieve higher quantum logic gate fidelities and approach the requirements for fault tolerant quantum computation. We thus expect DD to play a significant role in various quantum algorithms in the NISQ era. 

\begin{acknowledgments}
We thank Bibek Pokharel and Matthew Kowalsky for assistance with data collection scripts for the IBMQE devices.  We also acknowledge many useful and insightful discussions with Abhinav Kandala, Douglas T. McClure, Haimeng Zhang, Nicholas Ezzell, Humberto Munoz Bauza, Evgeny Mozgunov and Razieh Mohseninia. This material is based upon work supported by the National Science Foundation the Quantum Leap Big Idea under Grant No.~OMA-1936388. We acknowledge the use of IBM Quantum services for this work. The views expressed are those of the authors, and do not reflect the official policy or position of IBM or the IBM Quantum team. We acknowledge the access to advanced services provided by the IBM Quantum Researchers Program.
\end{acknowledgments}

\appendix

\begin{center}
\textbf{Supplementary Material}
\end{center}

\section{Data collection and analysis methodology}
\label{app:A}

\begin{figure}[t]
\subfigure{\includegraphics[width=1\linewidth]{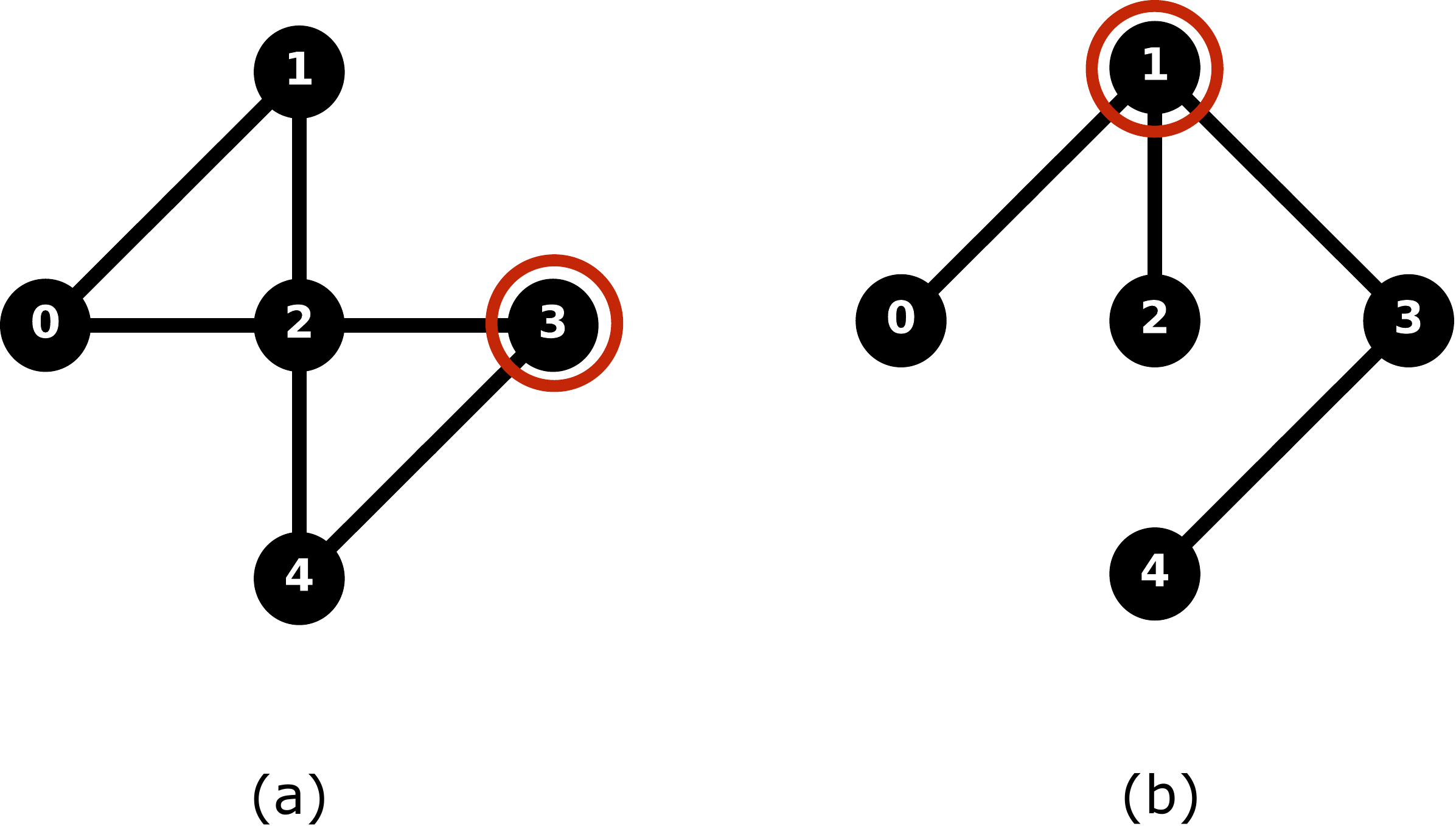}}
\caption{Schematics of the layout of the (a) Yorktown and (b) Ourense, Lima, and Quito devices. Thin circles indicate the main qubits used in our experiments. The other qubits are spectators. For Fig.~\ref{fig:main_2} of the main text we used qubit 1 (Q1) and qubit 3 (Q3) of the Quito as the control and target qubits, respectively.}  
\label{fig:1}
\end{figure}

The IBMQE devices used in this work are Ourense, Yorktown, Lima, and Quito, whose layout is shown schematically in Fig.~\ref{fig:1}. Figure~\ref{fig:main_1} of the main text uses qubit 1 (Q1) of Ourense and qubit 3 (Q3) of Yorktown as the main qubit. Figure~\ref{fig:main_2} of the main text uses qubit 1 (Q1) and qubit 3 (Q3) as the control and target qubits of the Quito processor, respectively. For Fig.~\ref{fig:4} below, we used Q1 of Lima and Q3 of Yorktown. All four devices are $5$-qubit processors consisting of superconducting transmon qubits.  Various calibration details and hardware specifications relevant to the qubits and gates used in this work are provided in Table~\ref{table1} and Table~\ref{table2}. Ourense and Yorktown data was acquired on 1/18/21 and 1/19/21, respectively. Quito data for Fig.~\ref{fig:main_2} of the main text was acquired on 11/30/2021.

\begin{table}[b]
\begin{adjustbox}{width=1\columnwidth,center}
\centering
\begin{tabular}{ |p{3cm}||p{1.7 cm}|p{1.7 cm}|p{1.7 cm}|p{1.7 cm} | }
 \hline
 Processor & Ourense  & Yorktown & Yorktown & Lima \\
 \hline
 Date accessed   & 01/18/2021   & 01/19/2021 & 07/07/2021   & 06/30/2021\\
 \hline
 Q0 &&&&\\
 Qubit freq. (GHz) & 4.8203 & 5.2828 & 5.2823 &  5.0298\\
$T_1\;(\mu \rm s)$  & 117.7   &  38.0 & 54.7  & 125.9\\
$T_2\;(\mu \rm s)$   & 79.4 & 23.1 &  22.6 &  134.0\\
sx gate error [$10^{-2}$] & 0.0310  & 0.1371 & 0.0204 & 0.0304\\
sx gate length\;(ns) & 35.556 & 35.556 & 35.556 & 35.556 \\
readout error [$10^{-2}$] & 1.60  & 2.92   & 10.99 & 2.48\\
\hline
Q1 &&&&\\
 Qubit freq. (GHz) & 4.8902 & 5.2476 & 5.2475 &  5.1276\\
$T_1\;(\mu \rm s)$  & 96.3   &  52.4 & 64.5  & 79.0\\
$T_2\;(\mu \rm s)$   & 29.6 & 23.2 &  27.4 &  140.3\\
sx gate error [$10^{-2}$] & 0.0368  & 0.1563 & 0.0954 & 0.0205\\
sx gate length\;(ns) & 35.556 & 35.556 & 35.556 & 35.556 \\
readout error [$10^{-2}$] & 3.35  & 3.00   & 33.7 & 1.83\\
\hline
Q2 &&&&\\
 Qubit freq. (GHz) & 4.7166 & 5.0335 & 5.0334 &  5.2474\\
$T_1\;(\mu \rm s)$  & 117.1   & 63.1 & 70.3  & 135.1\\
$T_2\;(\mu \rm s)$   & 114.5 & 87.6 &  36.9 &  174.0\\
sx gate error [$10^{-2}$] & 0.0668  & 0.0464 & 0.0522 & 0.0500\\
sx gate length\;(ns) & 35.556 & 35.556 & 35.556 & 35.556 \\
readout error [$10^{-2}$] & 1.76  & 7.68   & 9.72 & 1.55\\
\hline
Q3 &&&&\\
 Qubit freq. (GHz) & 4.7891 & 5.2923 & 5.2920 &  5.3034\\
$T_1\;(\mu \rm s)$  & 138.4   &  59.3 & 60.0  & 74.9\\
$T_2\;(\mu \rm s)$   & 106.7 & 43.8 &  28.2 &  48.0\\
sx gate error [$10^{-2}$] & 0.0374  & 0.0388 & 0.0537 & 0.0599\\
sx gate length\;(ns) & 35.556 & 35.556 & 35.556 & 35.556 \\
readout error [$10^{-2}$] & 3.48  & 5.54   & 3.03 & 10.21\\
\hline
Q4 &&&&\\
 Qubit freq. (GHz) & 5.0238 & 5.0785 & 5.0784 &  5.092\\
$T_1\;(\mu \rm s)$  & 110.2   &  47.0 & 56.7  & 21.1\\
$T_2\;(\mu \rm s)$   & 33.8 & 32.0 &  39.3 &  22.0\\
sx gate error [$10^{-2}$] & 0.0495  & 0.0703 & 0.0592 & 0.0802\\
sx gate length\;(ns) & 35.556 & 35.556 & 35.556 & 35.556 \\
readout error [$10^{-2}$] & 4.69  & 2.94   & 5.34 & 5.37\\
 \hline
\end{tabular}
\end{adjustbox}
\caption{Specifications of the Ourense, Yorktown, and Lima devices along with the access dates of our experiments. The sx ($\sqrt{\sigma^x}$) gate forms the basis of all the single qubit gates and any single qubit gate of the form $U3(\theta, \phi, \lambda)$ is composed of two sx and three rz$(\lambda) = {\rm exp}(-i\frac{\lambda}{2}\sigma^z)$ gates (which are error-free and take zero time, as they correspond to frame updates).}
\label{table1}
\end{table}

For each dataset in Fig.~\ref{fig:main_1} of the main text, we generated a series of circuits (one for each point on the corresponding curve), all of which were sent to the processor together in one job. We ensured that for each plot, all the jobs (corresponding to various curves) were sent consecutively and within the same calibration cycle in order to avoid charge-noise dependent fluctuations and variations in key features over different calibration cycles. We generated $70$ points on each curve  (note that we show only half of these points to avoid overcrowding) and thus the number of circuits sent in one job was $70$ (the maximum allowed number is $75$) and each circuit was repeated $8192$ times. The qubits were measured in the $Z$ basis and counts for both measurement outcomes ($0/1$) were recorded in ``dictionaries".  We define the empirical fidelity as the number of favorable measurement outcomes (of $0$) to the total number of experiments ($8192$). Error bars were then generated using the standard bootstrapping procedure, where we resample (with replacement) counts out of the experimental counts dictionary and create several new dictionaries. The final fidelity and error bars are obtained by calculating the mean and standard deviations over the fidelities of these new resampled dictionaries. Using $10$ such resampled dictionaries of the counts was enough to give sufficiently small error bars. We report the final fidelity with $2\sigma$ error bars, corresponding to $95\%$ confidence intervals. 

In Fig.~\ref{fig:main_1} of the main text all the single qubit gates---including the $X$ and $Y$ gates which are part of the DD sequences---are decomposed in terms of two ${\rm sx}$ ($\sqrt{\sigma^x}$) gates and and three virtual ${\rm rz}(\lambda)={\rm exp}(-i\sigma^z\lambda/2)$ gates. The latter are error-free and take zero time, as they correspond to frame updates.

\begin{figure}[t]
\subfigure{\includegraphics[width=1.\linewidth]{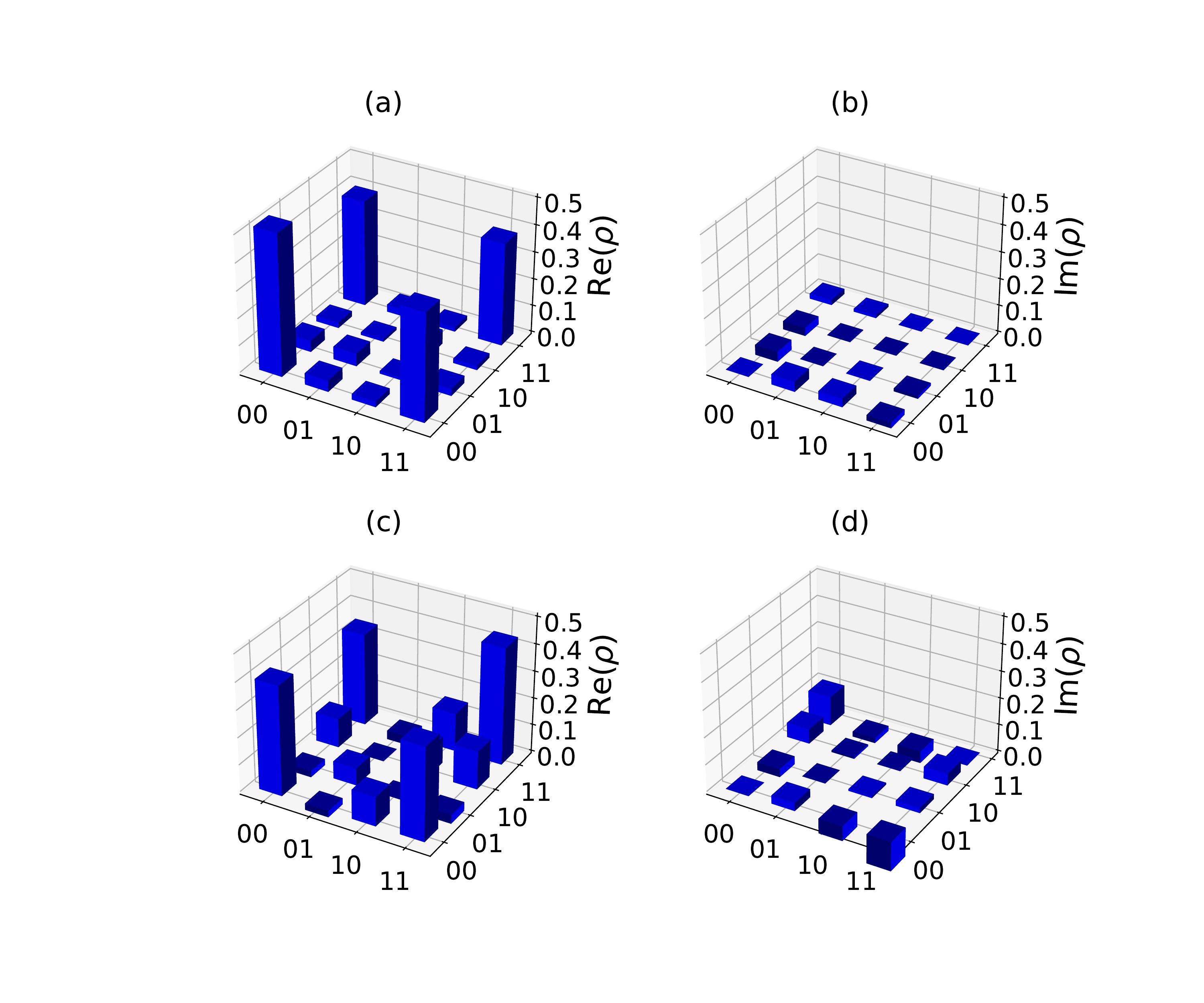}}
\caption{Quantum state tomography measurements after applying 1 CNOT gate [(a) and (b)] and 7 CNOT gates [(c) and (d)] on the $\ket{+,0}$ state of the control and target qubit. The fidelity with the ideal Bell state $(\ket{00}+\ket{11})/\sqrt{2}$ is $0.845$ after 1 CNOT gate and $0.761$ after 7 CNOT gates. Data acquired on 11/30/2021.}  
\label{fig:2}
\end{figure}

For Fig.~\ref{fig:main_2} of the main text, we chose two main qubits, i.e., a control and target qubit, leaving the remaining three qubits as spectators. We used Q1 and Q3 of Quito as the control and target qubits, such that two spectators are coupled to the control qubit and only one spectator is coupled to the target [see Fig.~\ref{fig:1}(b)]. We first applied a $R_{y}(\pi/2)$ gate to the control qubit, thus preparing the two main qubits in the state $\ket{+,0}$. We then applied a series of CNOT gates (control to target) and performed QST after every odd number of CNOT gates. The expected state is then the Bell state $(\ket{00}+\ket{11})/\sqrt{2})$. An example of a QST measurement is shown in Fig.~\ref{fig:2} where we show the measured density matrix after applying a single CNOT gate and after applying 7 CNOT gates. We then repeated the same experiment with DD incorporated into the circuits and present the comparison between the two cases in Fig.~\ref{fig:main_2}(b) of the main text. We explored different types of DD sequences for the spectator qubits coupled to both the control qubit and target qubit (Sec.~\ref{app:C}~b). In Fig.~\ref{fig:main_2}(b) of the main text, all the DD pulses used consist of only one ${\rm x}$ $(\sigma^{\rm x})$ gate and two virtual ${\rm rz}(\lambda)$ gates. Since both ${\rm sx}$ and ${\rm x}$ gates are available as calibrated backend gates on the IBMQE processors and are of equal length ($35.556$ ns), using ${\rm x}$ instead of ${\rm sx}$ gates as a basis gate for DD pulses in Fig.~\ref{fig:main_2}(b) of the main text enables fitting more DD pulses in parallel to the CNOT gates.

\begin{table*}[b]
\begin{adjustbox}{width=1.8\columnwidth,center}
\centering
\begin{tabular}{ |p{3cm}||p{1.7 cm}|p{1.7 cm}|p{1.7 cm}|p{1.7 cm}|p{1.7 cm}|p{1.7 cm}|p{1.7 cm}|}

 \hline
 Date accessed   &  12/30/2021 & 12/31/2021 & 01/01/2022 & 01/02/2022 & 01/07/2022 & 01/08/2022 & 11/30/2021\\
 \hline
 
 Q0 &&&&&&&\\
 Qubit freq. (GHz) &  5.3006 & 5.3006 & 5.3006  & 5.3006   & 5.3006  & 5.3006 &  5.3006\\
$T_1\;(\mu \rm s)$   & 55.8 & 84.6  &  96.7 &  79.8  &  114.7 &  95.2 & 85.7\\
$T_2\;(\mu \rm s)$    &  87.3 & 137.4 &  77.1 &  178.6  & 118.2  & 112.7 &  149.6\\
sx gate error [$10^{-2}$] & 0.0336 & 0.0309 & 0.0363  & 0.0325   &  0.0370 &  0.0257  & 0.0275\\
sx gate length\;(ns)  & 35.556 & 35.556 & 35.556 & 35.556 & 35.556& 35.556 & 35.556 \\
x gate error [$10^{-2}$]  & 0.0336 & 0.0309 & 0.0363  &  0.0325  &  0.0370 &  0.0257  & 0.0275\\
x gate length\;(ns)  & 35.556 & 35.556 & 35.556 & 35.556 & 35.556 & 35.556 & 35.556 \\
readout error [$10^{-2}$]   & 5.71 & 4.72 &  4.51 &  3.75  &  3.40 &  3.11  & 4.31\\
\hline

Q1 &&&&&&&\\
 Qubit freq. (GHz)  & 5.0806 & 5.0806  &  5.0806 &   5.0806  &  5.0806 &  5.0806  &  5.0806\\
$T_1\;(\mu \rm s)$   & 129.2  &  114.4  & 89.4  &  58.6  &  131.7 &  106.3  & 191.1\\
$T_2\;(\mu \rm s)$    & 122.4 &  98.2 &  59.8 &  59.9  &  140.9 &  154.0  &  82.1\\
sx gate error [$10^{-2}$]  & 0.04605 &  0.0597  &  0.0566 & 0.0283   &  0.0653 &  0.0247  & 0.0860\\
sx gate length\;(ns) & 35.556 & 35.556 & 35.556 & 35.556 & 35.556& 35.556 & 35.556 \\
x gate error [$10^{-2}$]   & 0.04605 & 0.0597   &  0.0566 &  0.0283  &  0.0653 &  0.0247  & 0.0860\\
x gate length\;(ns) & 35.556 & 35.556 & 35.556 & 35.556 & 35.556 & 35.556 & 35.556 \\
readout error [$10^{-2}$]   & 2.80 &  3.83  & 3.80  &  1.54  &  3.00 &  2.30  & 7.91\\
\hline

Q2 &&&&&&&\\
 Qubit freq. (GHz) & 5.3222 &  5.3222 & 5.3222  & 5.3222   &  5.3221 &  5.3222  &  5.3221\\
$T_1\;(\mu \rm s)$  & 90.0 &  87.8  &  76.9 &  120.9  &  94.2 &  114.2  & 77.0\\
$T_2\;(\mu \rm s)$    & 143.9 &  170.5  &  129.2 &  158.8  &  26.6 &  121.9  &  103.3\\
sx gate error [$10^{-2}$]   & 0.0370 & 0.0333  &  0.0452 &  0.0495  &  0.0371 & 0.1668 & 0.0352\\
sx gate length\;(ns) & 35.556 & 35.556 & 35.556 & 35.556 & 35.556& 35.556 & 35.556 \\
x gate error [$10^{-2}$]  & 0.0370 &   0.0333 &  0.0452 &  0.0495  & 0.0371  &   0.1668 & 0.0352\\
x gate length\;(ns)  & 35.556 & 35.556 & 35.556 & 35.556 & 35.556 & 35.556 & 35.556 \\
readout error [$10^{-2}$]   & 2.23 &  2.20 &  2.19 &  1.78  &  5.77  &  4.46  & 4.58\\
\hline

Q3 &&&&&&&\\
 Qubit freq. (GHz)  & 5.1636 &  5.1636 &  5.1636&   5.1636 &   5.1636 &  5.1636  &  5.1636\\
$T_1\;(\mu \rm s)$    & 125.1 & 81.7  &  112.7 &  112.2  &  110.5 &  113.3  & 111.3\\
$T_2\;(\mu \rm s)$   &  21.4 &  21.4 &  21.4 &  21.4  &  21.7 &   21.7 &  10.3\\
sx gate error [$10^{-2}$]   & 0.0257 & 0.0278  & 0.0299  & 0.0260  & 0.0238  &  0.0241  & 0.0288\\
sx gate length\;(ns)  & 35.556 & 35.556 & 35.556 & 35.556 & 35.556& 35.556 & 35.556 \\
x gate error [$10^{-2}$]  & 0.0257 &  0.0278 &  0.0299 & 0.0260 &  0.0238 &   0.0241 & 0.0288\\
x gate length\;(ns) & 35.556 & 35.556 & 35.556 & 35.556 & 35.556 & 35.556 & 35.556 \\
readout error [$10^{-2}$]   &  2.60 &  2.04 &  2.52 & 2.24  & 2.60 &  2.44  & 2.73\\
\hline

Q4 &&&&&&&\\
 Qubit freq. (GHz)  & 5.0524 &  5.0524 &  5.0524 &  5.0524 &  5.0524 &  5.0523  &  5.0524\\
$T_1\;(\mu \rm s)$    & 145.2 &  106.1 & 117.4  & 104.9  &  87.5 &  97.4  & 104.8\\
$T_2\;(\mu \rm s)$   & 171.5 &  187.9 & 113.4  &  160.2 &  110.6 & 180.7   &  203.5\\
sx gate error [$10^{-2}$]   &  0.0412 & 0.0291  &  0.0381 & 0.0418  & 0.0396  &  0.0299  & 0.0405\\
sx gate length\;(ns) & 35.556 & 35.556 & 35.556 & 35.556 & 35.556& 35.556 & 35.556 \\
x gate error [$10^{-2}$]  & 0.0412 & 0.0291  & 0.0381  & 0.0418  &  0.0396 &  0.0299  & 0.0405\\
x gate length\;(ns) & 35.556 & 35.556 & 35.556 & 35.556 & 35.556 & 35.556 & 35.556 \\
readout error [$10^{-2}$]   & 2.23 &  2.15 & 2.28  & 2.01  &  2.19 &  1.87  & 2.19\\
 \hline
 
 CNOT (Q1$\rightarrow$ Q3) &&&&&&&\\
 gate length\;(ns) &&&&&&& 334.222\\
gate error [$10^{-2}$] &&&&&&& 0.90\\ 
 
 \hline
\end{tabular}
\end{adjustbox}
\caption{Specifications of the Quito device along with the access dates of our experiments. Data for Fig.~\ref{fig:singleQubitAvg} was acquired over a period of six days from 12/30/2021 to 01/02/2022 and from 01/07/2022 to 01/08/2022 and the data for Fig.~2 of the main text was acquired on 11/30/2021.}
\label{table2}
\end{table*}

\begin{figure*}[t]
\subfigure{\includegraphics[width=.48\linewidth]{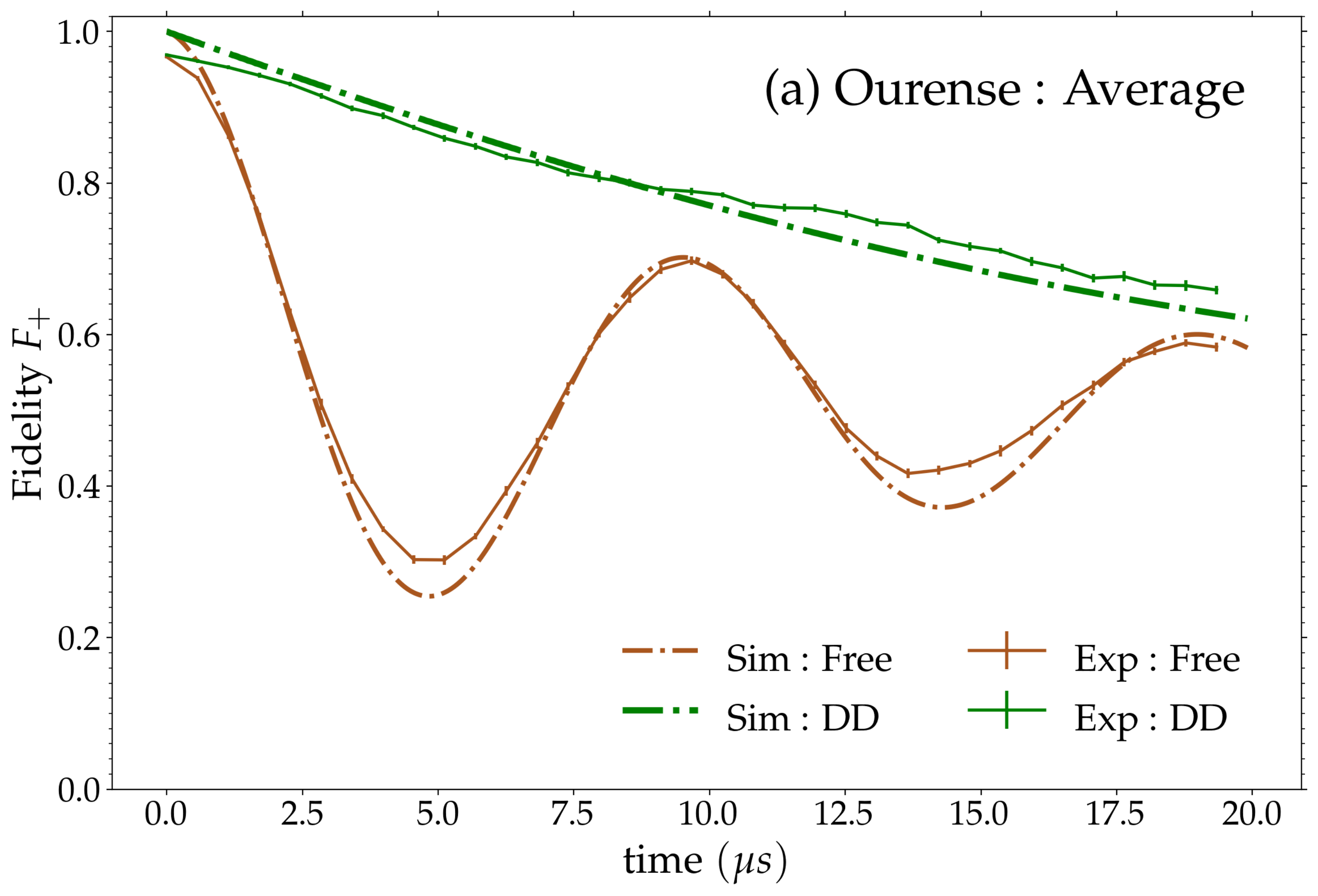}}
\subfigure{\includegraphics[width=.48\linewidth]{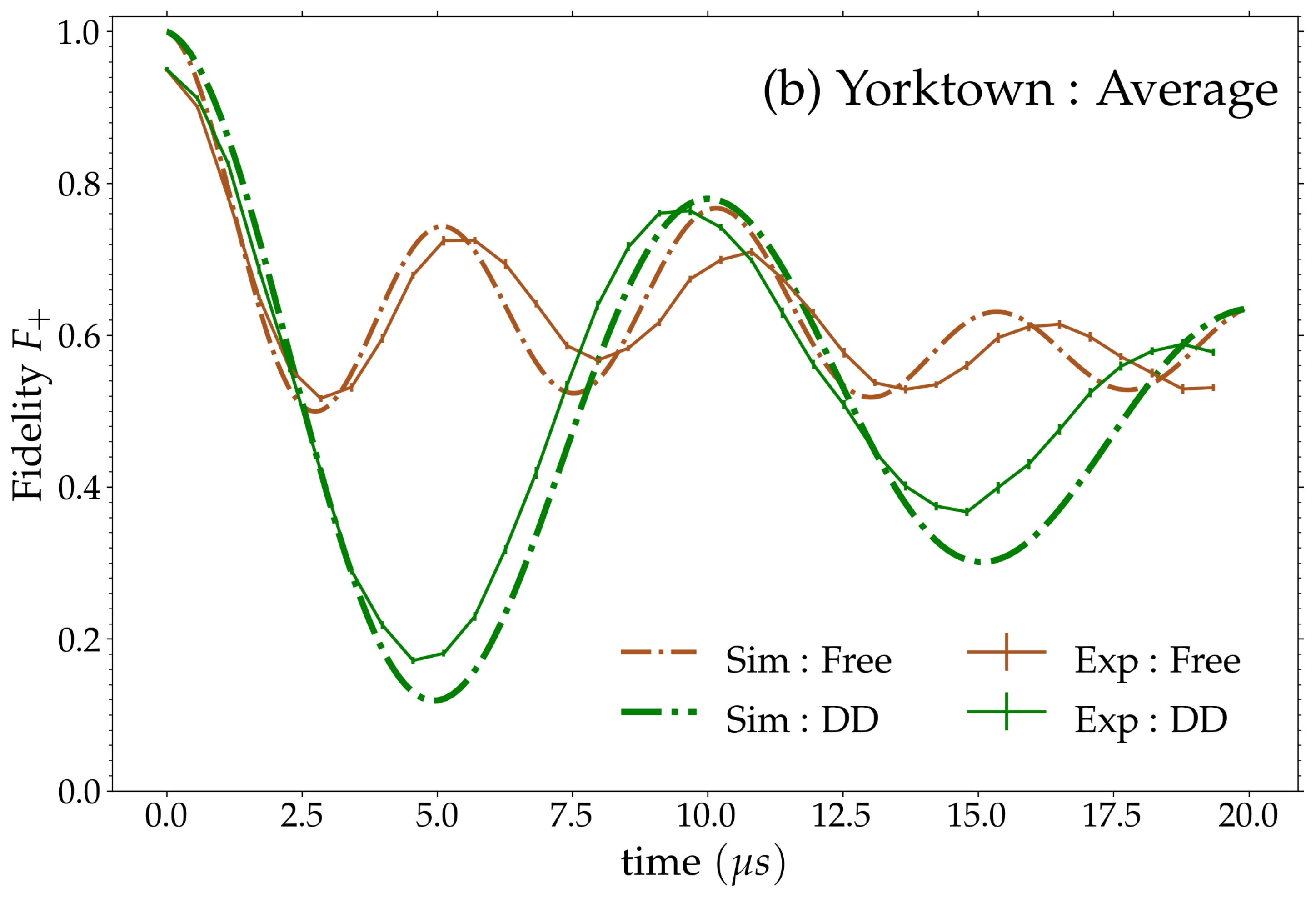}}
\caption{Results of Fig.~\ref{fig:main_1} of the main text averaged over all the three spectator qubit states for Ourense (a) and Yorktown (b). Note  that the simulations did not account for state preparation and measurement errors.}
\label{fig:3}
\end{figure*}

\begin{figure*}[t]
\centering
\includegraphics[width=.75\linewidth]{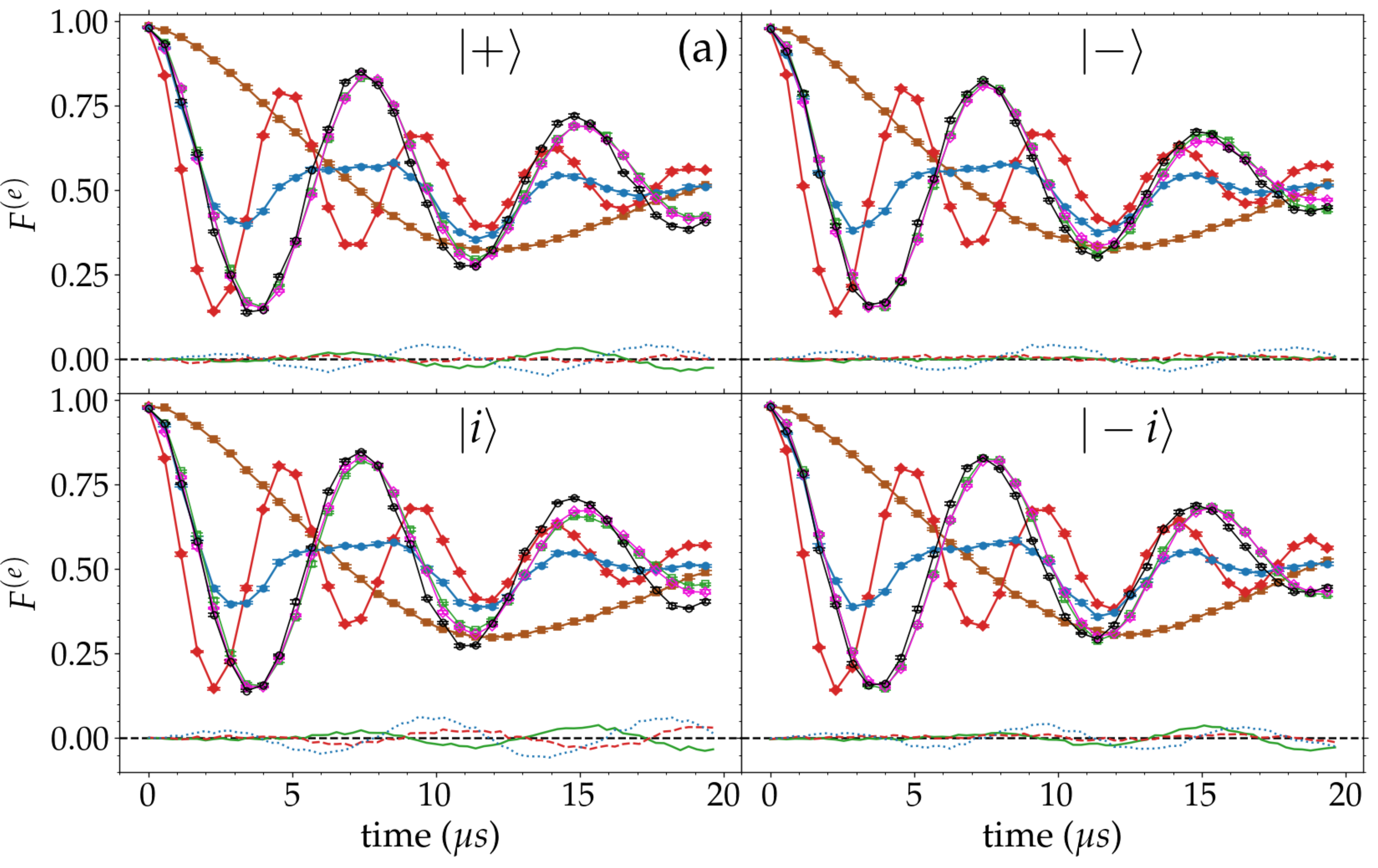}
\includegraphics[width=.75\linewidth]{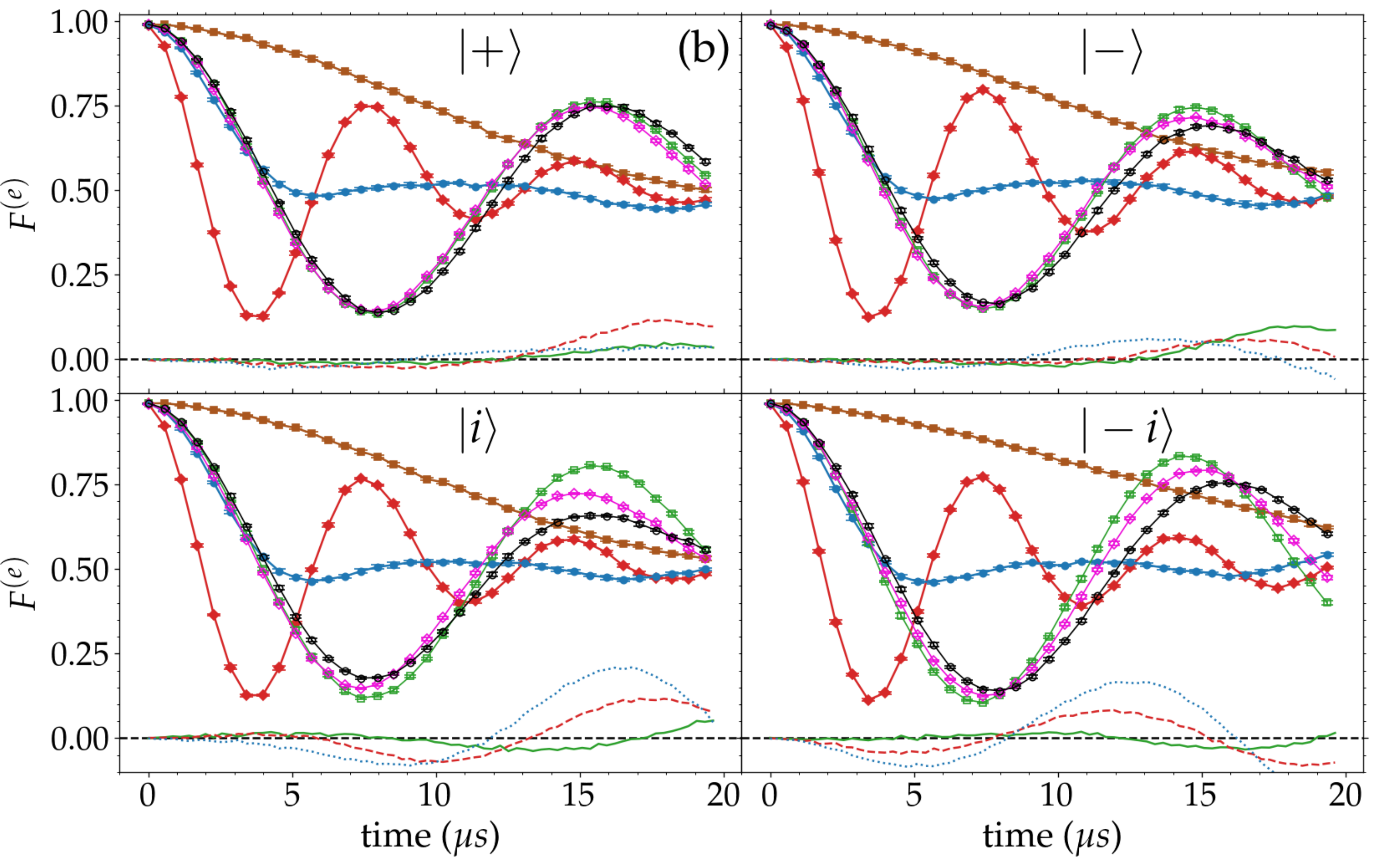}
\includegraphics[width=.6\linewidth]{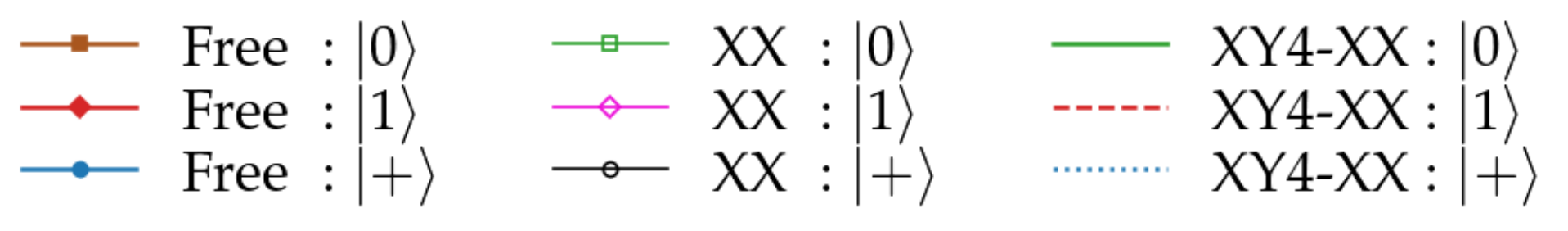}
\caption{Experimental results for different initial states $\{\ket{\pm}=(\ket{0}\pm\ket{1})/\sqrt{2},\ket{\pm i}=(\ket{0}\pm i\ket{1})/\sqrt{2}\}$ of the main qubit. The spectator qubit states are $\{\ket{0},\ket{1},\ket{+}\}$, as denoted in the legend. XY4-XX denotes the fidelity difference between the XY4 and XX sequences. (a) Yorktown results averaged over 5 runs. Data acquired on 7/7/21. (b) Lima results averaged over 4 runs. Data acquired on 6/30/21.}
\label{fig:4}
\end{figure*}

\begin{figure*}
\centering
\includegraphics[width=\linewidth]{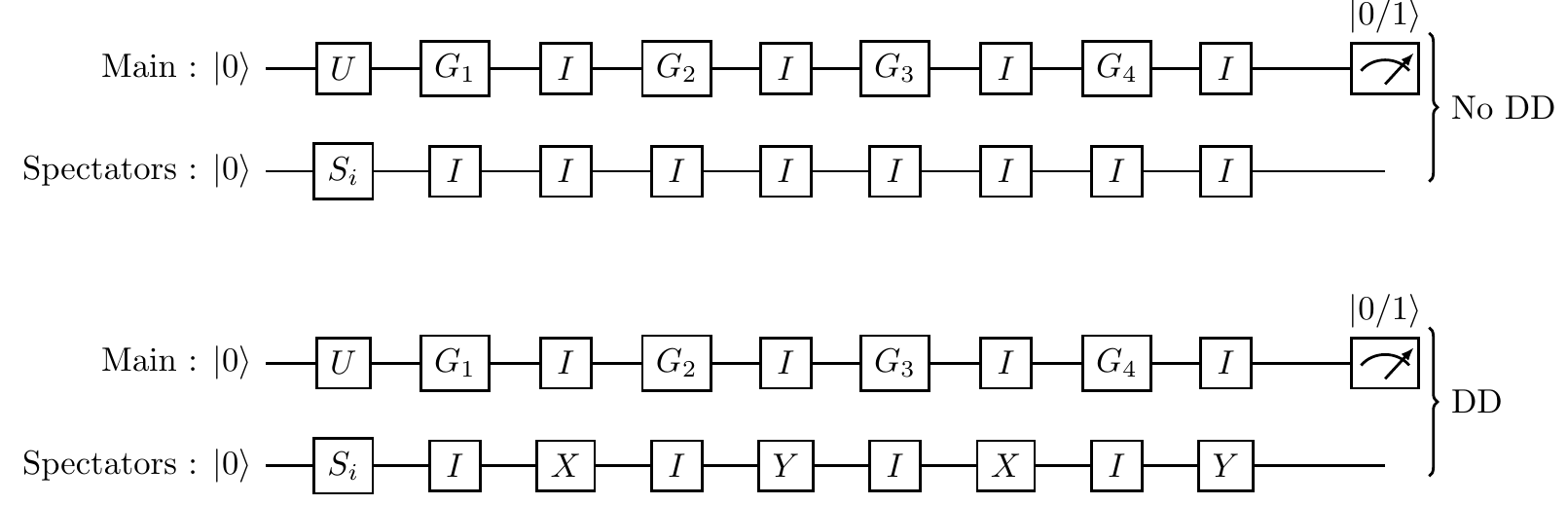}
\caption{Circuit diagram for single qubit gate experiments. The top circuit shows a sequence of random gates of length 4 chosen from the set $G=\{R_x(\pm\pi/8), R_x(\pm\pi/4) ,R_y(\pm\pi/8),R_y(\pm\pi/4)\}$ of 8 single qubit gates applied to the main qubit, and a sequence of Identity operations applied to all the spectator qubits. $U$ represents the gate applied to prepare any predefined initial state on the main qubit. In Fig.~\ref{fig:singleQubitAvg}, $U=R_{y}(\pi/2)$ which prepares a $|+\rangle$ state on the main qubit. $S_i$ represents the gate applied to prepare  $i=|0\rangle$, $|1\rangle$ and $|+\rangle$ on all the spectator qubits, where $S_{|0\rangle}=I$ (Identity). In the bottom circuit, the XY4 sequence is applied to the spectator qubits in the gaps between the gates applied to the main qubit.}
\label{fig-ckt}
\end{figure*}

\section{Additional experimental results}
\label{app:C}

\subsubsection{State protection experiments}
\label{app:C1}

In Fig.~\ref{fig:3} we present the same results as in Fig.~\ref{fig:main_1} of the main text, but averaged over all three spectator qubit initial states for each of the two IBMQE processors. This shows clearly that the envelope of the DD-protected evolution decays more slowly than that of the free evolution, in both cases. This figure also highlights the quantitative match between our numerical model and the experimental results.

In Fig.~\ref{fig:4} we present results for the Yorktown and Lima processors, for the main qubit initial states $\ket{\pm},\ket{\pm i}$, and for the spectator qubit states $\{\ket{0},\ket{1},\ket{+}\}$. These results are for free evolution and DD-protected evolution under the pure-X sequence (denoted XX in the figure). In addition, we compare the XY4 and pure-X sequences, by showing their fidelity differences. It is clear that the Lima processor is calibrated similarly to Yorktown, i.e., with the spectator qubits in the $\ket{0}$ state. In all cases the pure-X sequence removes the differences between the spectator initial states, as can be seen by the near overlap of the three XX curves in each subfigure (for $t>10$\;$\mu$s, small differences appear under DD). For both processors the fidelity differences between the XY4 and pure-X sequences are negligible for sufficiently short times; the difference is more noticeable for the Lima processor, where at longer times XY4 appears to be somewhat better (though it is hard to be certain since the advantage might have returned to pure-X had we been able to collect data for even longer times). We attribute this to stronger relaxation compared to dephasing on the Lima spectator qubits relative to the other processors; this is explained in more detail in Sec.~\ref{app:F}.

\begin{figure}[t]
\centering
\includegraphics[width=\linewidth]{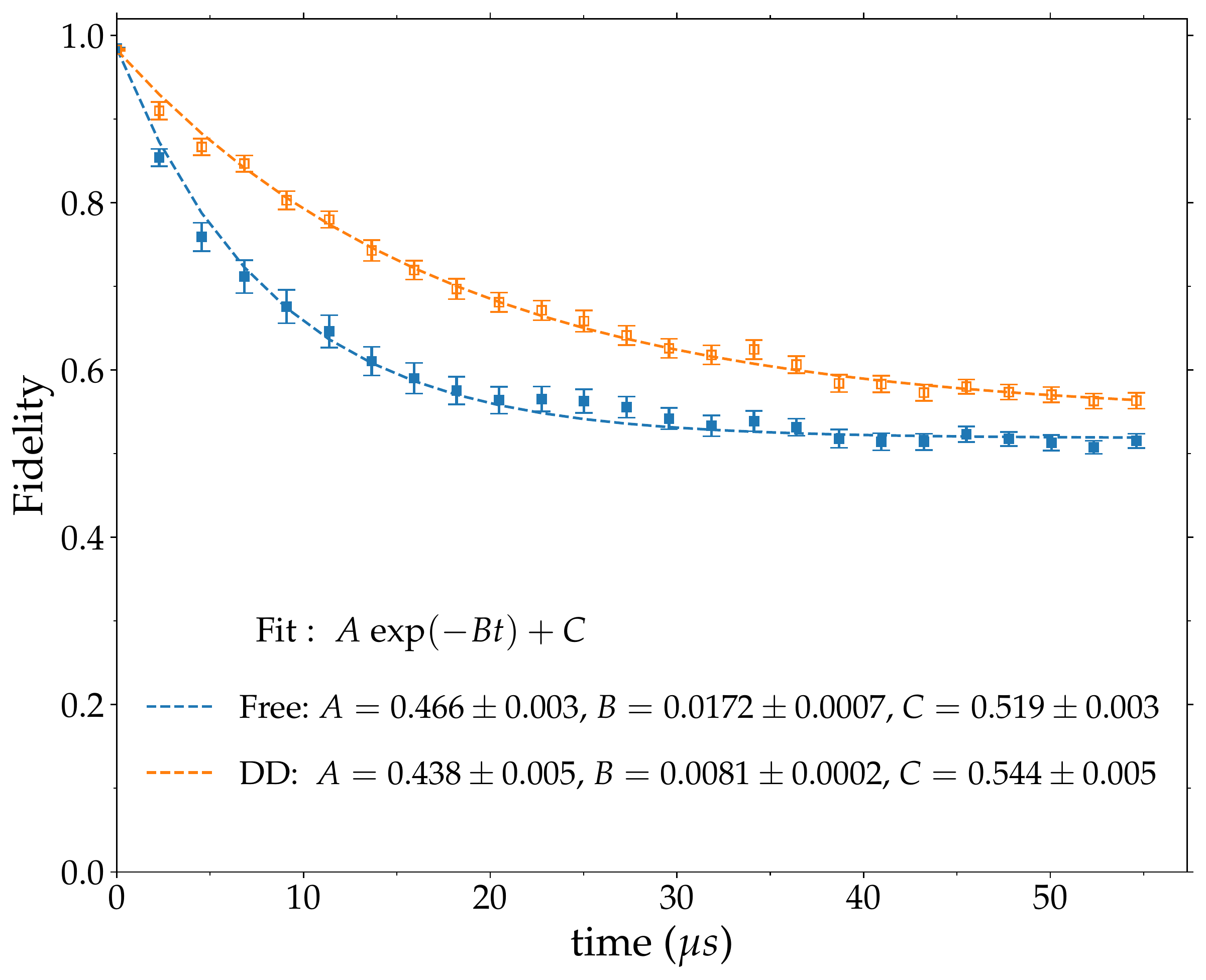}
\caption{Experimental fidelity results for random sequences of single qubit gates consisting of elements of the set $G=\{R_x(\pm\pi/8), R_x(\pm\pi/4), R_y(\pm\pi/8), R_y(\pm\pi/4)\}$, averaged over the three spectator states and $100$ different experimental runs, with the main qubit initialized in the $|+\rangle$ state. The fidelity is shown as a function of time, with and without DD applied to the spectator qubits. The exponential fit to both the free and the DDPGs case shows a clear improvement in the decay rate, by a factor of $0.0172/0.0081=2.12$. Error bars represent $2\sigma$ confidence intervals obtained by bootstrapping. Data was acquired over a period of six days from 12/30/2021 to 01/02/2022 and from 01/07/2022 to 01/08/2022. See Table~\ref{table2} for device parameters.}
\label{fig:singleQubitAvg}
\end{figure}

\begin{figure*}[t]
\centering
\subfigure[\ ]{\includegraphics[width=.48\linewidth]{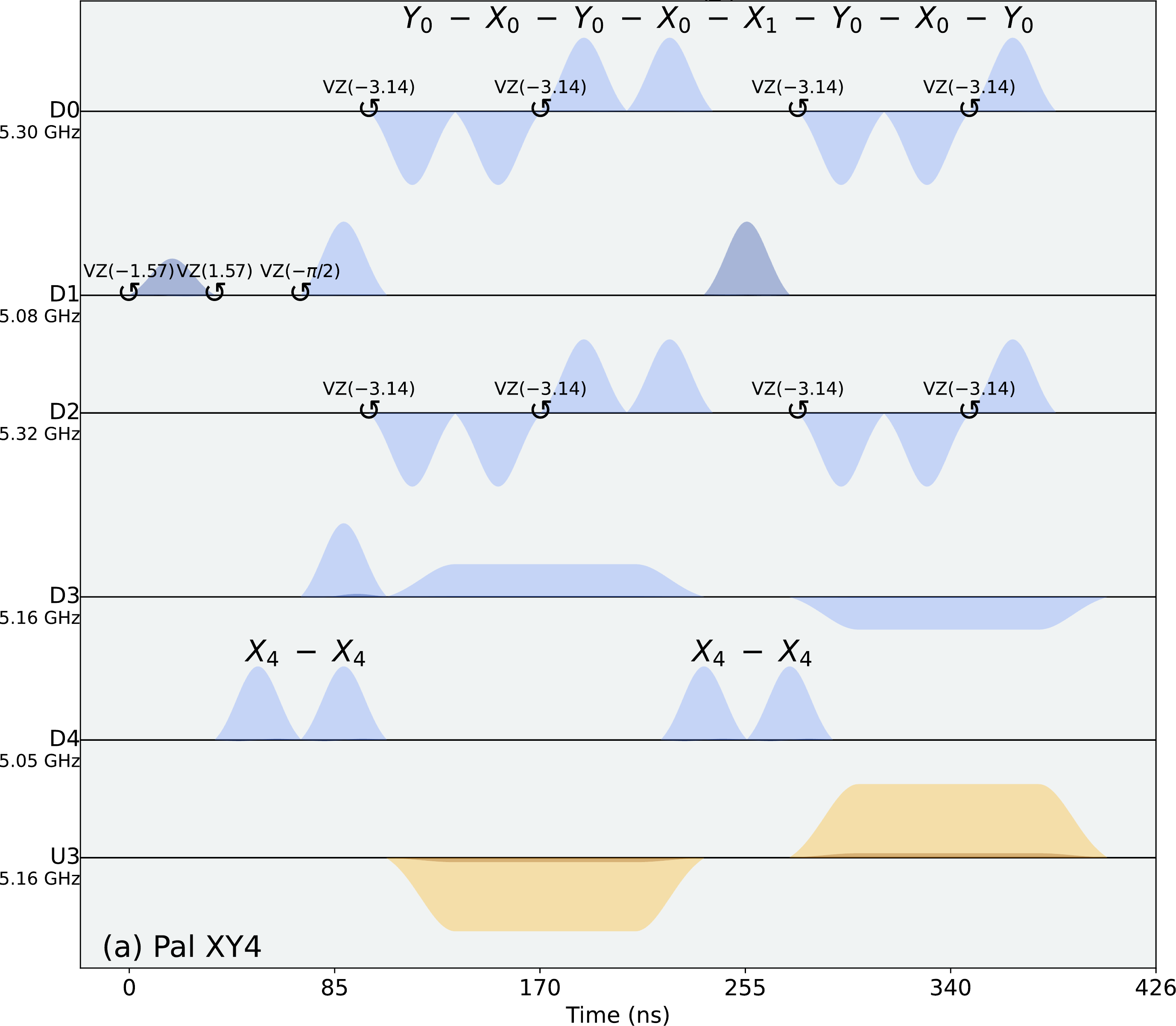}}
\subfigure[\ ]{\includegraphics[width=.48\linewidth]{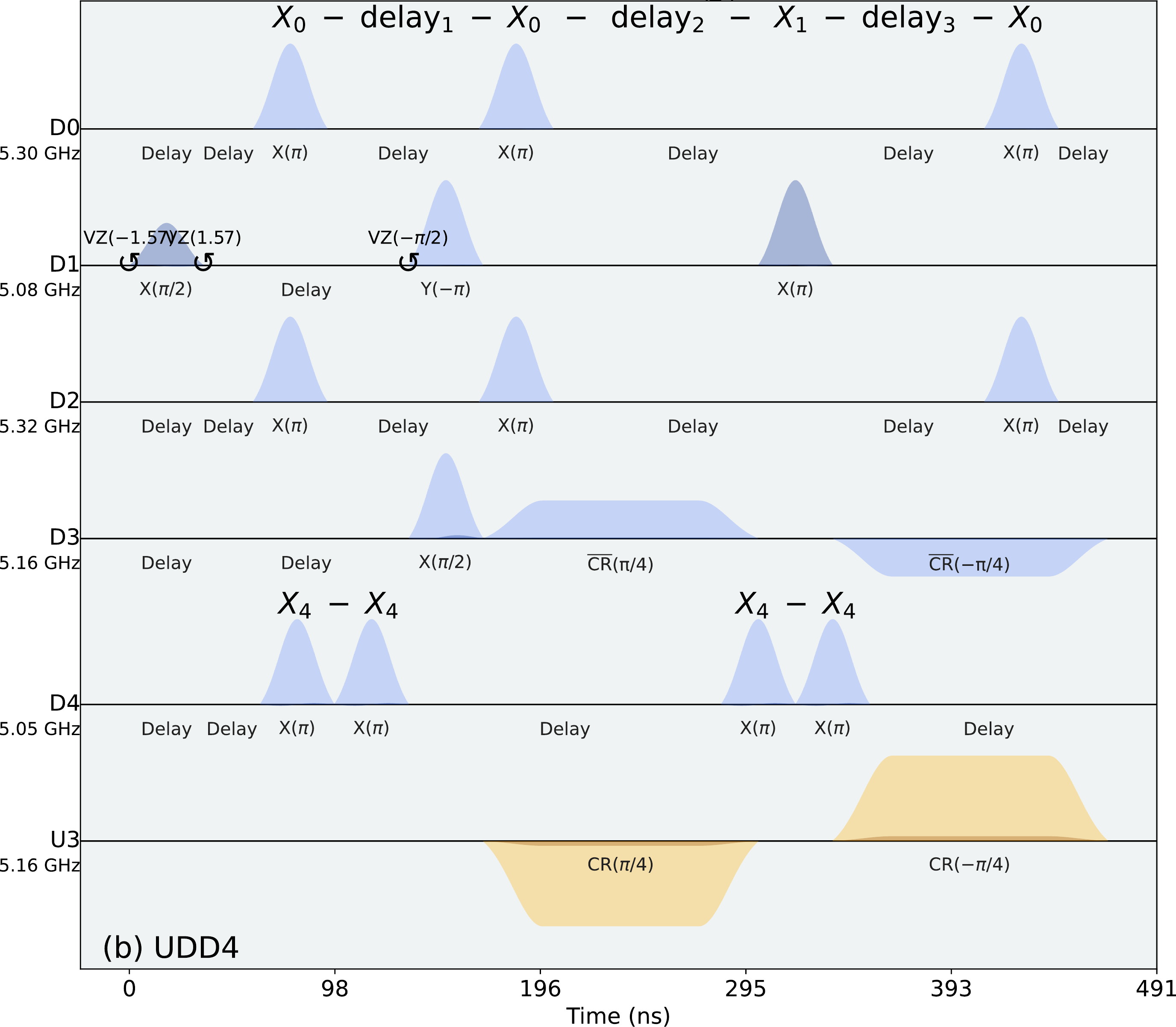}}
\caption{Pulse schedule diagram for the two-qubit experimental circuit, consisting of state preparation on drive channel D1 corresponding to qubit 1 (Q1) by applying an $R_y(\pi/2)$, followed by a CNOT gate between Q1 and Q3 (control and target, respectively). The CNOT gate consists of two single qubit gates on Q1 (D1) and a single qubit gate on Q3 (D3), two CR pulses on Q1 (D1) at the frequency of Q4 (here shown on a separate channel U3 but actually acting on the drive channel D1 corresponding to Q1), and two rotary pulses on Q3 (D3). The DD sequence used on target spectator qubit Q4 (D4) consists of two instances of pure-X DD before the first rotary pulse and in between the two CR pulses. The DD sequence on the control and spectator qubits Q0 (D0) and Q2 (D2) consists of two repetitions of XY4 (XYXY) in Fig.~2(a) of the main text, one repetition of palindromic XY4 in (a), and one repetition of UDD4 in (b). Note that one $X$ gate in each of the sequences applied to D0 and D2 was replaced by the $X$ gate on D1. These diagrams are obtained using schedule.draw() feature of qiskit~\cite{schedule-draw}.} 
\label{fig:schedule}
\end{figure*}

\begin{figure*}[t]
\centering
\subfigure[\ ]{\includegraphics[width=.32\linewidth]{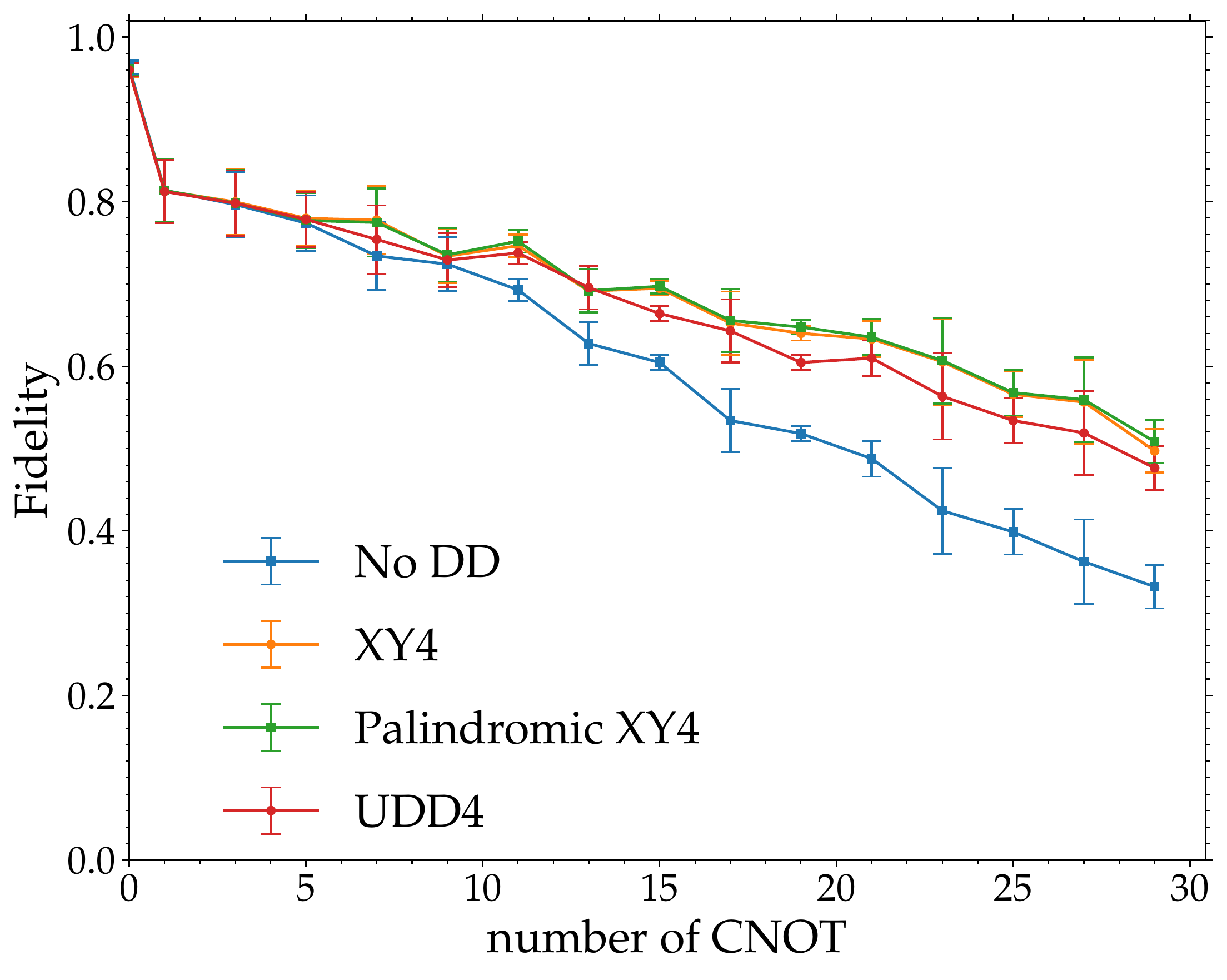}\label{fig:twoQubitDD}}
\subfigure[\ ]{\includegraphics[width=.32\linewidth]{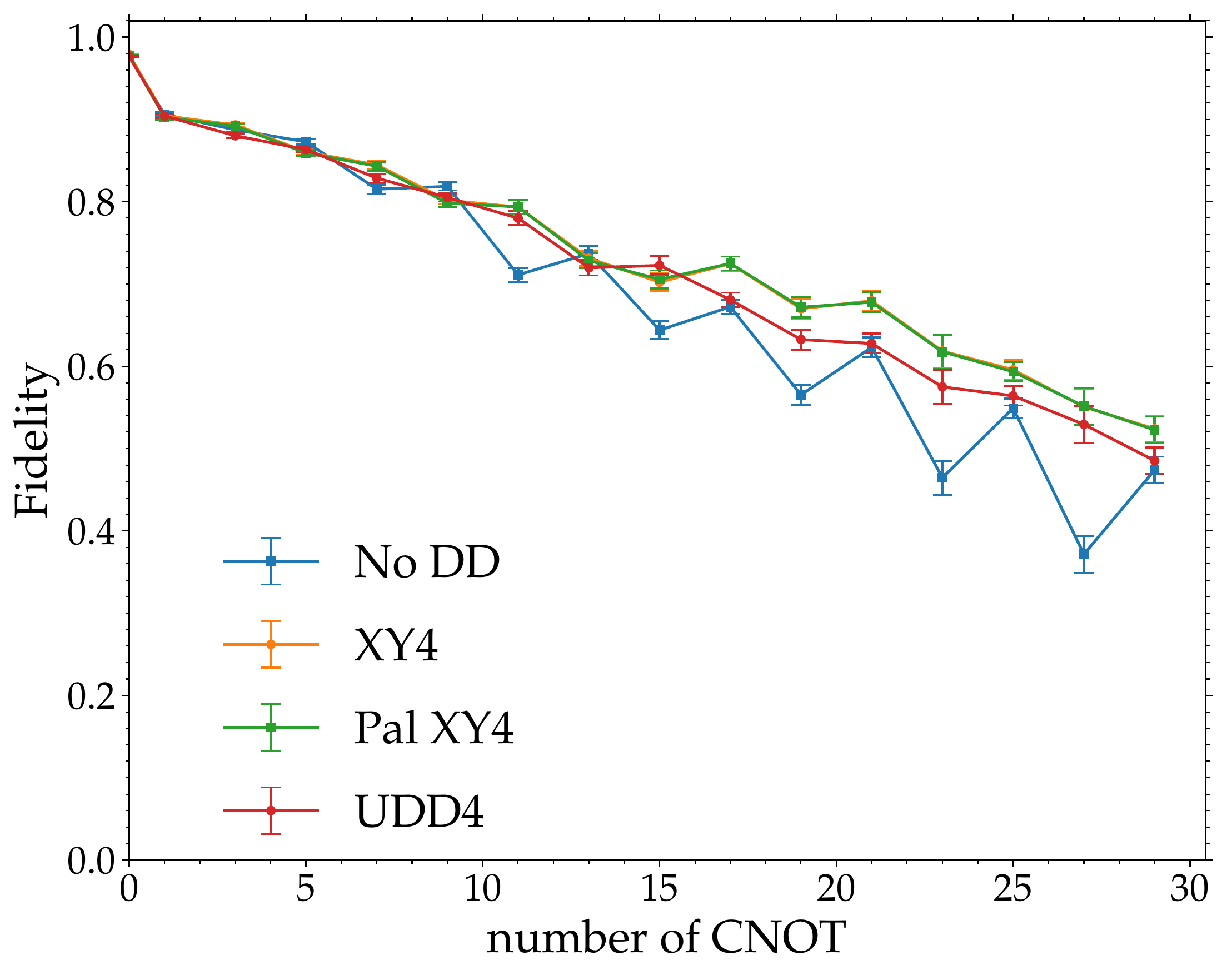}\label{fig:two-qubit-avg_b}}
\subfigure[\ ]{\includegraphics[width=.32\linewidth]{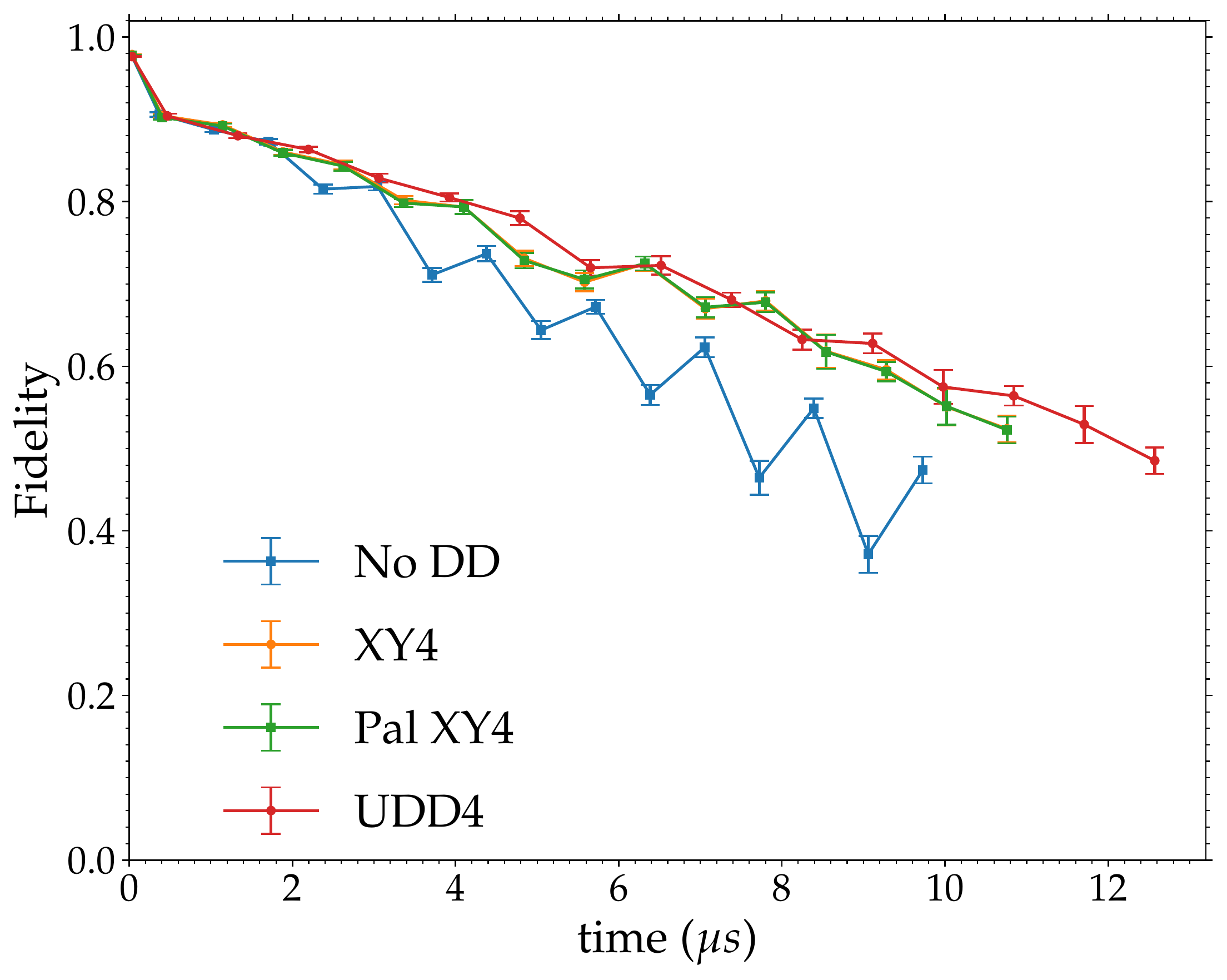}\label{fig:two-qubit-avg_a}}
\caption{Fidelity of standard and DD-protected CNOT gates, for different DD sequences. (a) This plot complements Fig.~\ref{fig:main_2}(c) in the main text, which shows the result as a function of actual time elapsed. The fidelity is shown here as a function of the number of CNOT gates applied to the $\ket{+,0}$ state of the control and target qubits (Q1 and Q3 of Quito) without DD and with three different DD sequences (XY4, palindromic XY4 and UDD4) applied to the control spectator qubits in parallel to the CNOT gate operation, averaged over five experimental runs. Since different DD sequences require a different delay between subsequent CNOT gates, we obtain a different total time for the same number of CNOT gates, as in Fig.~\ref{fig:main_2}(c) in the main text. Data was acquired on 11/30/2021. For (b) and (c) data was acquired over a period of four different calibration cycles from 01/18/2022 to 01/22/2022. (b) Same as (a), but averaged over the four different calibration cycles. (c) Same as (b), showing instead the fidelity as a function of time, as in Fig.2(c) in the main text.} 
\label{fig:twoQubitDD-all}
\end{figure*}

\subsubsection{Single-qubit gates experiments}
\label{app:C2}

In our single-qubit gate experiments, we chose Q1 of Quito as the main qubit and generated a series of circuits consisting of random sequences of gates of varying length, where each of the gates is taken from the set $G=\{R_x(\pm\pi/8), R_x(\pm\pi/4) ,R_y(\pm\pi/8),R_y(\pm\pi/4)\}$ of 8 single qubit gates, where $R_{x/y}(\theta)$ represents a rotation about the $x/y$-axis by an angle $\theta$. We performed quantum  state tomography (QST) to construct the density matrix at the end of each of the circuits. This involved repeating the same experiment thrice, measuring each time in a different Pauli basis. The  results for each basis were again given as counts. We compared the density matrix thus obtained with the expected state to calculate the fidelity as a function of time or number of gates. We used the bootstrap method by resampling with replacement over the number of counts, as discussed above, to create the resampled density matrix and thus an average fidelity for each of the circuits. In the DD-protected gates (DDPGs) case, a DD sequence (XY4) was simultaneously applied to all the spectator qubits, in parallel with the gates we applied to the main qubit. The DD pulses occupy only the gap between each of the gates on main qubit; see the circuit diagram in Fig.~\ref{fig-ckt}.

For the results shown in Fig.~\ref{fig:singleQubitAvg}, we first prepare the main qubit in the $|+\rangle$ state, and apply the above sequences of random gates with identity operations (or delay) between any two consecutive gates, such that the distance between the center of any two consecutive gates is twice the total duration of the gates (see Fig.~\ref{fig-ckt}). We choose the same random circuit and repeat it for three different initial states of the spectator qubits: $\{\ket{0},\ket{1},\ket{+}\}$. 
We repeat the whole experiment 100 times, with each run having its own random sequence of gates, with and without DD on the spectators. Figure~\ref{fig:singleQubitAvg} shows the result of averaging over three spectator qubits states and 100 experimental runs (each with a different random gate sequence).  We observe a clear and statistically significant improvement in fidelity for the DDPGs case. In particular, the exponential decay rate decreases by more than a factor of $2$, from $(0.0172\pm 0.0007)$\,MHz without DD, to $(0.0081\pm 0.0002)$\,MHz with DD.

\subsubsection{Two-qubit gates experiments}
\label{app:CNOT}

We elaborate on the DD-protected CNOT gate results reported in the main text. Recall that we initialized the control and target qubits in the $\ket{+,0}$ state, followed by an odd number of applications of the CNOT gate. Ideally this should return the Bell state $(\ket{00}+\ket{11})/\sqrt{2}$ each time. We tried different combinations of DD sequences on the spectator qubits coupled to both the control qubit and target qubit, which we now refer to as control spectators and target spectators. We found that the effect of the DD pulses depends on whether they are applied to the control or target spectators. The Quito CNOT gate is based on the ``echo cross-resonance'' (CR) gate~\cite{Corcoles2013}, which consists of two CR pulses on the control qubit, two rotary pulses on the target qubit~\cite{Sundaresan2020}, and several single qubit gates. We found that applying DD pulses to the target spectators during the rotary pulses causes the fidelity to drop, whereas applying DD pulses to the control spectators during the CR pulses improves fidelity. The former is unsurprising, since the rotary pulses can be interpreted as being part of the DD sequence, and in this sense another overlapping DD pulse on another qubit can have an adverse effect. 
We therefore picked a DD sequence where the target spectator DD pulses are always separated from the target qubit pulses. We also avoided any simultaneous single qubit gates (e.g., $XX$) on the spectator qubits and the control or target qubit; as shown in Fig.~\ref{fig:schedule}(a) and (b), we integrated single-qubit pulses which are inherently part of the CNOT gate into the DD sequence. The reason is that $XX$ commutes with the $ZZ$ cross-talk we set out to suppress.

Given all the constraints on avoiding overlap, we are limited to only simple DD sequences on the target spectator qubit (indicated by D4 in Fig.~\ref{fig:schedule}). We added an additional delay before each CNOT gate to accommodate a pure-X sequence on D4 at the beginning and between the two CR rotary pulses, avoiding any direct overlap. This gave the best results for any DD sequence we tried on the target spectator. We explored different DD sequences for the control spectator qubit D0. These include two repetitions of XY4 (XYXY), one repetition of palindromic XY4 (YXYXXYXY; a sequence known to result in theory in higher order suppression than XY4~\cite{Ng:2011dn,Khodjasteh:2007zr}) and one repetition of the 4th order Uhrig DD sequence (UDD4), consisting of four $X$ pulses with non-uniform intervals (also known to yield higher order suppression in theory than XY4~\cite{Uhrig:2007qf}). Figure~\ref{fig:schedule} complements Fig.~\ref{fig:main_2}(a) in the main text and shows the pulse schedule diagram of preparation of the control qubit in the $\ket{+}$ state, followed by a single CNOT gate with palindromic XY4 (a) and UDD4 (b) applied to the control spectators (D0 and D2), and two instances of pure-X applied to the target spectator qubit (D4). 

In Figs.~\ref{fig:twoQubitDD} and~\ref{fig:two-qubit-avg_b}, we present the fidelity results as a function of the number of CNOT gates, as obtained with different DD sequences on the control spectator (always with two instances of pure-X applied to the target spectator qubit). This complements Fig.~\ref{fig:main_2}(c) of the main text, which shows the same as a function of total time elapsed. For all of the DD sequences presented here we observe a statistically significant fidelity improvement over the no DD case, that grows with the number of applied CNOT gates [Figs.~\ref{fig:twoQubitDD} and~\ref{fig:two-qubit-avg_b}], as well as with the actual elapsed time [Fig.~\ref{fig:two-qubit-avg_a}]. The different DD sequences (XY4, palindromic XY4, and UDD4) applied to Q0 and Q2 do not have a statistically significant effect.

To test the robustness of our DD scheme in enhancing the CNOT gate performance, we repeated the two-qubit experiments over four different calibration cycles of Quito (which usually takes place every 24 hours), and plot the averaged results from 46 different experimental runs in Figs.~\ref{fig:two-qubit-avg_b} and~\ref{fig:two-qubit-avg_a}. The effect of both coherent and incoherent noise channels changes gradually over a period of several hours, and with each new calibration cycle the fidelity of both the single and two qubit gates changes. After averaging over four different calibration cycles over a period of a week, Figs.~\ref{fig:two-qubit-avg_b} and~\ref{fig:two-qubit-avg_a} still show a significant improvement for our DD-protected CNOT gates, with the additional desirable effect of removing the fidelity oscillations present in the no DD (standard CNOT) case. This additional data strengthens our conclusion that DD provides a significant advantage that is robust across different calibration cycles.

\section{Open quantum system model}
\label{app:B}

Here we describe the noise model used and the procedure to numerically simulate the open system dynamics. We consider a system of $n$ coupled qubits with only the linear system-bath interaction given by
$H_{\rm SB} = \sum_{i=0}^{n-1} \sum_{\a\in\{x,y,z\}} g_{i\a} \s_i^\a \otimes B_{i\a}$ where $\s_{i}^{\a}$ and  $B_{i\a}$ represents the system and bath coupling operators and $g_{i\a}$ are coupling strengths. In our simulations, we included the main qubit and all the spectator qubits which are directly coupled to the main qubit. Therefore, our simulations of the Ourense device included four qubits (see Fig.~\ref{fig:1}), where Q1 is the main qubit, coupled to three spectator qubits), and similarly, we included only three qubits in the Yorktown simulations, where Q3 is the main qubit. 
The system Hamiltonian in this case is given by:
\begin{equation}
H_{\mathrm{S}}=-\sum_{i=0}^{n-1} \frac{\omega_{q_{i}}}{2} Z_{i}+\sum_{j >i=0}^{n-1} J_{ij} Z_{i} Z_{j}\ .\label{eq:multiqubit_Ham}
\end{equation}
We again move to a rotating frame defined by the number operator $\hat{N}$ given by 
\begin{equation}
\hat{N} = \sum_{i_0,..,i_{n-1}\in \{0,1\}}(i_0+\cdots +i_{n-1})\ketb{i_0 \dots i_{n-1}}{i_0\dots i_{n-1}}\ ,
\label{eq:number_operator}
\end{equation}
and solve the Redfield (or TCL2) master equation~\cite{Breuer:book,chen2020hoqst} for the rotated system Hamiltonian. Introducing a superindex \{$\rm m$\} instead of \{$i\a$\}, we define the standard bath correlation function:
\begin{equation}
    C_{\rm mn}(t-\tau) = g_{\rm m}g_{\rm n}\Tr\{U_{\rm B}(t-\tau)B_{\rm m}U^\dagger_{\rm B}(t-\tau) B_{\rm n}\rho_{\rm B}\}\ ,
\end{equation}
where $U_{\rm B}(t)=e^{-iH_{\rm B}t}$ is the unitary generated by the pure-bath Hamiltonian $H_{\rm B}$, and the reference state $\rho_{\rm B}$ is the Gibbs state of $H_{\rm B}$:
\begin{equation}
    \rho_{\rm B}=e^{-\beta H_{\rm B}}/\Tr  \big(e^{-\beta H_{\rm B}}\big) \ ,
\end{equation}
where $\beta = 1/T$ is the inverse temperature.
Assuming the bath operators $B_\mathrm{m}$ and $B_\mathrm{n}$ are uncorrelated, i.e., $C_{\mathrm{mn}}(t)=C_{\mathrm{nm}}(t)=\delta_{\mathrm{mn}}C_{\mathrm{n}}(t)$, the Redfield equation is
\begin{equation}
    \label{eq:hybrid_redfield}
    \frac{\partial\rho_\mathrm{S}}{\partial t}= -i[H_{\rm S}, \rho_\mathrm{S}] + \mathcal{L}(\rho_\mathrm{S}) \ ,
\end{equation}
where $\mathcal{L}$ is the Redfield Liouvillian
\begin{equation}
    \mathcal{L}(\rho_\mathrm{S}) = -\sum_{\rm m} [A_{\rm m}, \Lambda_{\rm m}(t)\rho_\mathrm{S}(t)] + \text{h.c.} \ ,
\end{equation}
and
\begin{equation}\label{eq:redfield_lambda}
    \Lambda_{\rm m}(t) = \int_0^{t} C_{\rm m}(t-\tau)U_{\rm S}(t, \tau)A_{\rm m}U_{\rm S}^\dagger(t, \tau) \mathrm{d}\tau \ ,
\end{equation}
where $U_{\rm S}(t)=e^{-iH_{\rm S}t}$ is the unitary operator generated by the system Hamiltonian $H_{\rm S}$. Note that the $A_{\rm m}$ used in our simulations are the coupling operators defined in the rotating frame of the number operator $\hat{N}$ [Eq.~\eqref{eq:number_operator}].

We choose the bath to be Ohmic, which means that the noise spectrum,
\begin{equation}
    \gamma_{\rm m}(\omega) = \int_{-\infty}^{\infty} C_{\rm m}(\tau) e^{i\omega\tau} \mathrm{d}\tau \ ,
\end{equation}
has the following form
\begin{equation}
    \gamma_{\rm m}(\omega) =2 \pi \eta g_{\rm m}^{2} \frac{\omega \mathrm{e}^{-|\omega| / \omega_{\mathrm{m}}}}{1-\mathrm{e}^{-\beta \omega}} \ ,
    \label{eq-Ohmic}
\end{equation}
where $\omega_{\rm m} = 2\pi f_{\rm m}$ is the cutoff frequency for bath operator $B_{\rm m}$, and $\eta$ is a positive
constant with dimensions of time squared that arises in the specification of the Ohmic spectral function.

We work in units such that $\hbar=1$ and assume the bath temperature $T=20\;{\rm mK}$. For all the simulations shown in Figs.~\ref{fig:main_1}(c) and (d) of the main text, we used 
\bes
\begin{align}
\eta &= 10^{-4}\;{\rm GHz}^{-2},\ f_{\rm m} =  2\;\text{GHz} \;\forall {\rm m} \\ 
g_{ i, \sigma^z} &= 0.1175 \;{\rm GHz}\\
g_{ i, \sigma^{x}}&=g_{ i, \sigma^{y}} = \begin{cases}
\frac{1}{2}g_{ i, \sigma^z} \text{ (Ourense)}\\
\frac{3}{4}g_{ i, \sigma^z} \text{ (Yorktown)}
\end{cases}
\forall i\ .
\end{align}
\ees  
For the Ourense device, the $J_{ij}$ are provided in the IBMQE device backend information~\cite{qiskit-backendproperties} and take the following values:
\beq
J_{01} = 25.48 {\rm\; KHz},\; J_{12} = 18.24 {\rm\; KHz},\;J_{13} = 8.77 {\rm\;KHz}.
\eeq 
The $J_{ij}$ are not provided for the Yorktown device; therefore we assume that all the non-zero $J_{ij}$ are equal and we extract $J_{ij} = 24.27 \;$KHz by tuning it to match the oscillations in Fig.~\ref{fig:main_1}(b) of the main text.

Finally, we remark that the values of the bath parameters reported here are chosen to provide a qualitative agreement with the experimental data. A more rigorous optimization scheme to fit the bath parameters in order to obtain a quantitative match between theory and simulations with multi-qubit systems will be the subject of a future publication. 

We use the open system model described here to obtain Fig.~\ref{fig:main_1}(c) and Fig.~\ref{fig:main_1}(d) of the main text. 
Figure~\ref{fig:main_1}(c) shows the simulation results for the $s={+}$ frame.
The effect of the bath is to induce an overall exponential decay envelope due to dephasing , as already
suggested by Eq.~(3a) of the main text. When DD is applied to the spectator qubit, the $ZZ$-induced oscillations are entirely suppressed, independently of the initial state of the spectator qubit,  as in the experimental data in Fig.~\ref{fig:main_1}(a) of the main text. 
Additionally, the DD sequence can be seen to suppress the coupling of the spectator qubit to the bath, in the sense that the maximum amplitude for the $\ket{1}$ and $\ket{+}$ spectator qubit states (at $10\mu$s) is lower in the free evolution case than in the DD case. The small but noticeable difference between the DD-protected curves is due to the fact that the DD sequence only generates first order suppression. 

Figure~\ref{fig:main_1}(d) shows the simulation results for the $s={0}$ frame.
As already expected from Eq.~(3) of the main text, the oscillation period for the free evolution cases with the spectator qubit prepared in the $\ket{1}$ or $\ket{+}$ states is $2\pi/4J \approx 5\mu$s, while the $\ket{0}$ case exhibits no oscillations. 
These results are entirely consistent with the experimental data in Fig.~\ref{fig:main_1}(b) of the main text.
In the presence of DD pulses applied to the spectator qubit the three cases again collapse onto a single curve. However, this time oscillations persist with a frequency of $2J$. As we discuss in the main text, these are due to the presence of the uncanceled $Z_1$ term in $\tilde{U}' (2\tau)$, with $\o_{\rm d} = \o_{q_1} -2J$. Crucially, despite the dependence on $J$, this is a single-qubit effect, and the goal of suppressing an unwanted two-qubit term that would interfere with proper two-qubit gate operation has been accomplished.
Similar comments as in the $s=+$ frame apply to the effect of DD on suppressing the effect of coupling to the bath; the envelope amplitude of the $\ket{1}$ and $\ket{+}$ cases is higher in the presence of DD. The $\ket{0}$ case is not helped by DD, since in this case there is no relaxation of the spectator qubit.

\section{Circuit model description of $ZZ$ coupling and its implications on rotating frame analysis}
\label{app:D}

\begin{figure}[t!]
		\subfigure{\includegraphics[width=1\linewidth]{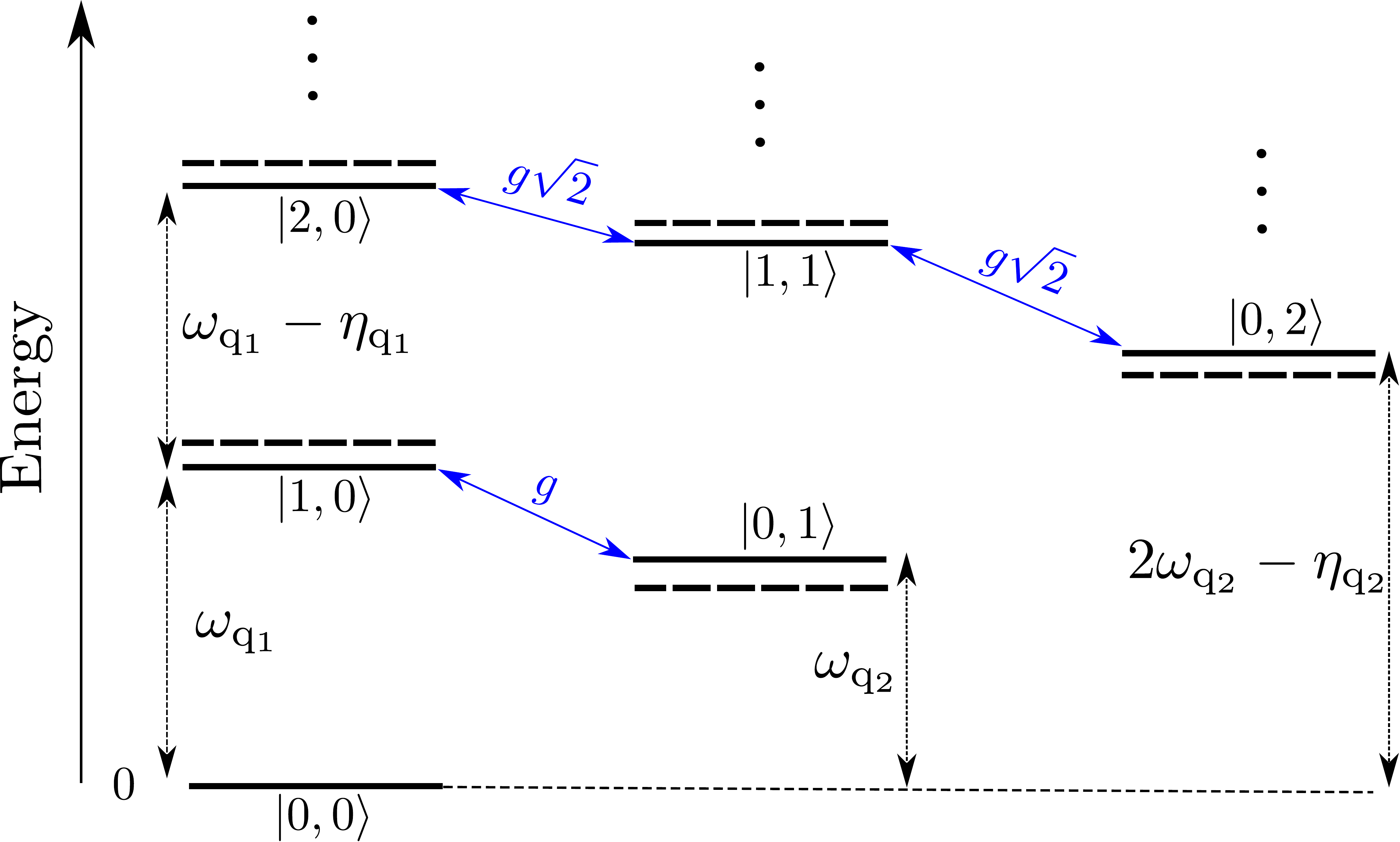}}
		\caption{Energy level diagram of two coupled transmons with qubit frequencies $\o_{q_1}$ and $\o_{q_2}$ and anharmonicities $\eta_{q_2}$ and $\eta_{q_1}$, coupled linearly with strength $g$. The solid lines represent the bare energy levels and dashed lines represent the eigenlevels. $\ket{k,l}$ represents levels $k$ and $l$ in the main and spectator transmons, respectively. Only $6$ levels of the infinite-dimensional Hilbert space formed by both transmons are shown.}
\label{energy level}
\end{figure}

In practice, transmons are not perfect two-level systems but anharmonic oscillators consisting of multiple levels. Here we show how $ZZ$ coupling arises in the multi-level model of coupled transmons, and verify its implications on free evolution, which was discussed in the main text based on a simplified model of two-level systems. 

The energy level diagram in the lab frame of two capacitively coupled transmons with an always-on coupling strength $g$ is shown in Fig.~\ref{energy level}~\cite{Tripathi:2019vb}. 
Because of the coupling $g$, levels $\ket{1,0}$ and $\ket{0,1}$ are repelled and form the dashed lines representing the eigenstates $\ket{\overline{1,0}}$ and $\ket{\overline{0,1}}$ with energies $E_{\ket{\overline{1,0}}} = \o_{q_1}+g^2/\Delta$ and $E_{\ket{\overline{0,1}}} = \o_{q_2}-g^2/\Delta$ where $\Delta = \o_{q_1}-\o_{q_2}$ and we have assumed $g/\Delta \ll 1$. Therefore, the main qubit eigenfrequency when the spectator qubit is in $\ket{0}$ is:
\begin{equation}
    \o_{\rm eig}^0 = E_{\ket{\overline{1,0}}}-E_{\ket{\overline{0,0}}} = \o_{q_1} + \frac{g^2}{\Delta}\label{omega_m_0}\ .
\end{equation}
Similarly, $\ket{1,1}$ is pushed downward by $\ket{2,0}$ and upward by $\ket{0,2}$. Therefore, we have:
\begin{equation}
    E_{|\overline{1,1}\rangle} = \o_{q_1} +\o_{q_2} -\frac{2g^{2}}{\Delta-\eta} +  \frac{2g^{2}}{\Delta+\eta}\ ,
\end{equation}
where we have assumed that $\eta_{q_1} = \eta_{q_2} = \eta$. Thus, the main qubit eigenfrequency when the spectator qubit is in $\ket{1}$ is:
\bes
\begin{align}
    \o_{\rm eig}^1 &= E_{|\overline{1,1}\rangle}-E_{\ket{\overline{0,1}}} \\
    &= \o_{q_1} -\frac{2g^{2}}{\Delta-\eta} +  \frac{2g^{2}}{\Delta+\eta} + \frac{g^2}{\Delta}\ .\label{omega_m_1}
\end{align}
\ees

Now the $ZZ$ coupling strength can be defined as~\cite{Tripathi:2019vb}:
\bes
\begin{align}
    2J &= \omega_{zz} = \frac{\o_{\rm eig}^1 - \o_{\rm eig}^0}{2} \\
    &=   \frac{g^{2}}{\Delta+\eta} - \frac{g^{2}}{\Delta-\eta} \ .
\end{align}
\ees
Using Eqs.~\eqref{omega_m_0} and~\eqref{omega_m_1}, we can also define the eigenfrequency of the main qubit when the spectator qubit is in $\ket{+}$ and is given as
\bes
\begin{align}
    \tilde{\o}_{\rm eig}^+ &= \frac{\o_{\rm eig}^0 + \o_{\rm eig}^1}{2}\\
    &= \o_{q_1} + \frac{g^2}{\Delta} - \frac{g^2}{\Delta - \eta} + \frac{g^2}{\Delta + \eta}\ .
\end{align}
\ees
Note that unlike the two-level system case discussed in the main text, this eigenfrequency ($\tilde{\o}_{\rm eig}^+$) is not same as the bare qubit frequency $\o_{q_1}$, and this is one sense in which the two-level system model is oversimplified. 
We now choose the drive frequency as $\omega_{\rm d} = \tilde{\o}_{\rm eig}^+$ and move into a rotating frame about the number operator $\hat{N}= \sum_{k,l}(k+l)\ketb{k,l}{k,l}$. With this choice, the eigenfrequencies of the main qubit for the spectator qubit in $\ket{0}$ and $\ket{1}$ are, respectively:
\bes
\begin{align}
\tilde{\o}_{q_1}^{0} &= \o_{\rm eig}^0 - \omega_{\rm d} =  -\omega_{zz}\\
\tilde{\o}_{q_1}^{1} &= \o_{\rm eig}^1 - \omega_{\rm d} = \omega_{zz}\ .
\end{align}
\ees
Therefore, in the $s = +$ frame, we have oscillations with frequency $\omega_{zz} = 2J$ irrespective of the state of the spectator qubit, which is exactly what we showed in the main text for the simplified two-level system model. We can similarly verify the results for the $s=0$ and $s=1$ frames.

\section{Analysis of free evolution under dephasing in the $s=+$ and $s=0$ frames}
\label{app:E}

All the calculations reported in this section are supported by a Mathematica package~\url{https://www.dropbox.com/s/ajsac2xjhj405op/free-evolution-calcs.nb?dl=0}, which can be used to reproduce and test all the claims made below.

Consider the phenomenological Lindbladian 
\bes
\label{eq:Lind-s}
\begin{align}
\mathcal{L}^s &= -i[\tilde{H}_{\rm S}^s,\cdot]+\sum_{\a} \g_\a (L_\a\cdot L_\a^\dag -\frac{1}{2}\{L_\a^\dag L_\a,\cdot\})\\
L_1 &= ZI\ ,\ L_2 = IZ\ ,\ L_3 = ZZ\ 
\end{align}
\ees 
specified in the main text, along with the Hamiltonians $\tilde{H}_{\rm S}^s$ given in Eq.~(2) for the rotating frames $s=+$ and $s=0$. The solution of 
\beq
\dot{\r} = \mathcal{L}^s\r
\label{eq:Lind}
\eeq 
is the joint density matrix $\r(t)$ of the main and spectator qubits. We are interested in the main qubit state $\r^s_{+s'}(t) = \Tr_{\rm spec}[\r(t)]$, where the partial trace is over the spectator qubit, given the initial state $\r(0) = \ketb{+s'}{+s'}$, with $s'\in\{+,0,1\}$ denoting the three spectator qubit initial states.

Since the Lindbladian $\mathcal{L}^s$ involves only diagonal operators, solving Eq.~\eqref{eq:Lind} is straightforward. Letting $\g = \g_1+\g_3$, we find, for the Hamiltonian~[Eq.~(2b)]:
\bes
\label{eq:rsys1}
\begin{align}
\r^+_{++}(t) &= \frac{1}{2}\left(
\begin{array}{cc}
 1 & e^{-2 \g t}  \cos (2 J t) \\
  e^{-2 \g t} \cos (2 J t)  & 1 \\
\end{array}
\right)\\
\r^+_{+0}(t) &= \frac{1}{2}\begin{pmatrix}
                 1 & e^{-2 \g t} e^{2iJt} \\
                 e^{-2 \g t} e^{-2iJt} & 1
               \end{pmatrix} = (\r^+_{+1}(t))^*\ ,
\end{align}
\ees
and for the Hamiltonian~[Eq.~(2a)]:
\bes
\label{eq:rsys2}
\begin{align}
\r^0_{++}(t) &=  \frac{1}{2}\begin{pmatrix}
                 1 & e^{-2 \g t} \frac{1+e^{-4iJt}}{2} \\
                 e^{-2 \g t} \frac{1+e^{4iJt}}{2} & 1
               \end{pmatrix} \\
\r^0_{+0}(t) &=  \frac{1}{2}\left(
\begin{array}{cc}
1 & e^{-2 \g t} \\
 e^{-2 \g t} & 1 \\
\end{array}
\right)\\
\r^0_{+1}(t) &= \frac{1}{2}\begin{pmatrix}
                 1 & e^{-2 \g t}e^{4iJt} \\
                 e^{-2 \g t} e^{-4iJt} & 1
               \end{pmatrix}\ .
\end{align}
\ees

We are interested in the probability of the main qubit remaining in the $\ket{+}$ state, which is given by 
\beq
p^s_{+s'}(t) = \bra{+}\r^s_{+s'}(t)\ket{+}\ .
\eeq 
Computing this quantity from Eqs.~\eqref{eq:rsys1} and~\eqref{eq:rsys2} directly yields Eq.~(3) from the main text.

Focusing on the $s=+$ case, we have also solved variations on Eq.~\eqref{eq:Lind-s} with additional Lindblad operators. Specifically, when including $\{L_4=XI,L_5=IX,L_6=XX\}$ in addition to $\{L_1,L_2,L_3\}$, with corresponding rates $\{\g_4,\g_5,\g_6\}$, we find:
\bes
\label{eq:px}
\begin{align}
\label{eq:pxa}
& p^{+}_{+s'}(t) = \frac{1}{2}\left[1+e^{-[2(\g_1+\g_3)+\g_4+\g_5]t}\times  \right.   \\ 
&\left. \left(\cos(2J_x't)+\frac{\g_4+\g_5}{2J'_x}\sin(2J'_x t) \right)\right] \quad \forall s'\in\{+,0,1\} \nonumber 
\\
2J'_x &\equiv \sqrt{4J^2-(\g_4+\g_5)^2}\ .
\end{align}
\ees
When, instead, we include $\{L_7=YI,L_8=IY,L_9=YY\}$ in addition to $\{L_1,L_2,L_3\}$, with corresponding rates $\{\g_7,\g_8,\g_9\}$ (setting $\{\g_4=\g_5=\g_6=0\}$), we find:
\bes
\label{eq:py}
\begin{align}
\label{eq:pya}
& p^{+}_{+s'}(t) = \frac{1}{2}\left[1+e^{-[2(\g_1+\g_3+\g_9)+\g_7+\g_8]t}\times \right.  \\
&\left. \left(\cos(2J_y't) +\frac{\g_8-\g_7}{2J'_y}\sin(2J'_y t) \right)\right] \quad \forall s'\in\{+,0,1\} \notag
\\
2J'_y &\equiv \sqrt{4J^2-(\g_8-\g_7)^2}\ .
\end{align}
\ees
Note that the $YY$ rate appears in the overall decay rate [$\g_9$ in Eq.~\eqref{eq:pya}], but the $XX$ decay rate does not [$\g_6$ is absent in Eq.~\eqref{eq:pxa}]. Also, note that the oscillation frequency in both models is modified by the single-qubit ($X$ or $Y$) rates, and is no longer simply $2J$, but rather $2J'_x$ or $2J'_y$. 

\begin{figure}[t!]
		\includegraphics[width=1\linewidth]{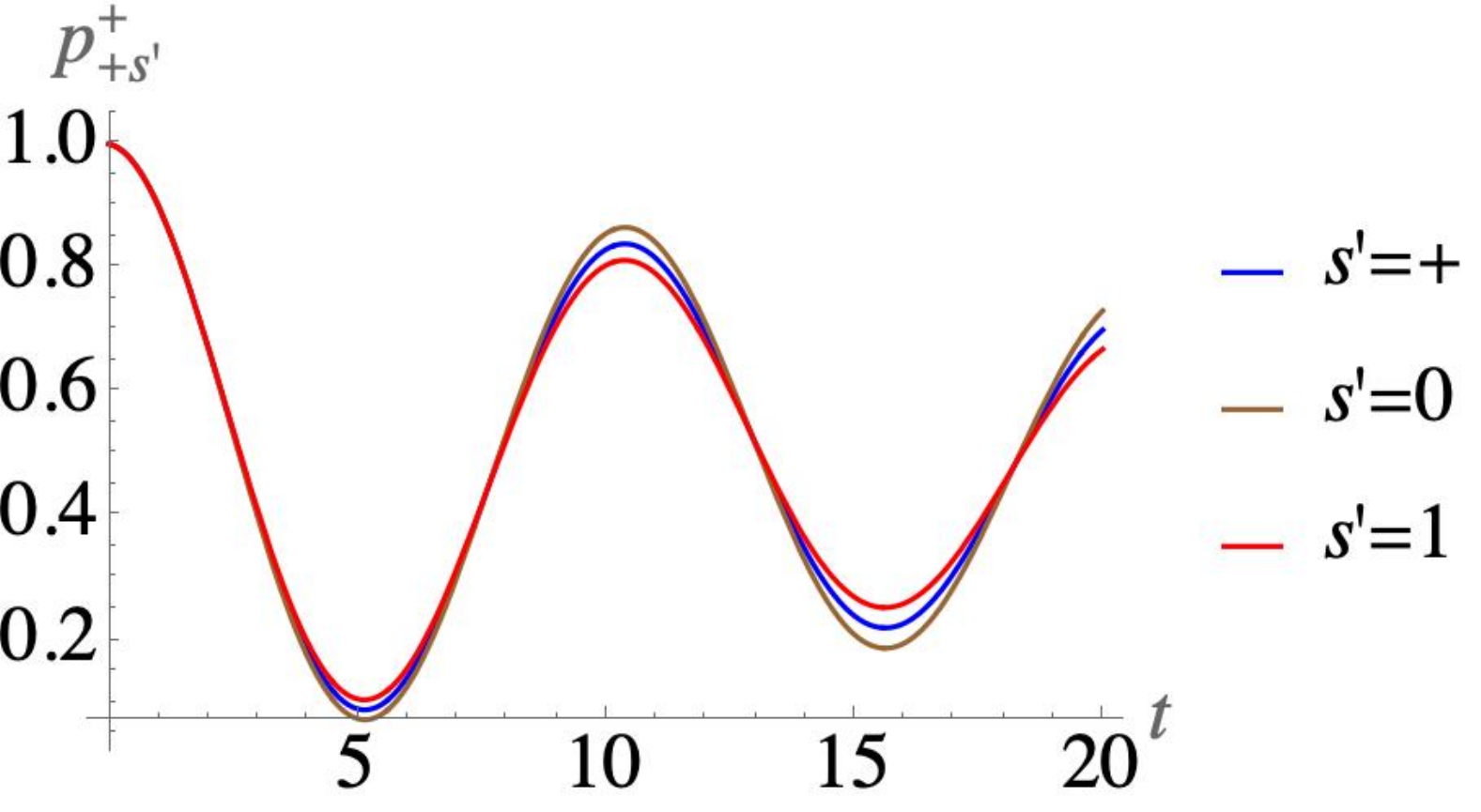}
		\caption{The fidelity expressions given in Eq.~\eqref{eq:Lind-Z+SE} for the following parameter values: $J=0.3, \g_1=0.01, \g_3=0.0025, \g_{10}=\g_{11}=\g_{12}=0.01$ (all in dimensionless units, as is $t$). The ordering of fidelity oscillation amplitudes is preserved for all parameter values we tried for this dephasing plus spontaneous emission model. Compare to Fig.~1(a) and (c) of the main text.}  
\label{fig:5}
\end{figure}

However, since neither Eq.~\eqref{eq:pxa} nor Eq.~\eqref{eq:pya} depends on $s'$, neither one of these two models correctly predicts the different amplitudes observed for the three different initial spectator states in Fig.~1(a) of the main text, so they do not appear to correctly describe the open system dynamics. If, instead we introduce spontaneous emission by including the lowering operator $\s^-=\ketb{0}{1}$ via the Lindblad operators $\{L_{10}=\s^-\ox I,L_{11}= I\ox \s^-,L_{12}=\s^-\ox \s^-\}$ (setting $\{\g_i=0\}_{i=4}^9$), we find that the symmetry between the different initial spectator states is broken:
\bes
\label{eq:Lind-Z+SE}
\begin{align}
p^{+}_{+0}(t) &= \frac{1}{2} \left(1+\cos (2 J t) e^{-(2 (\g_1+\g_3)+\g_{10}/2)t}\right)\\
p^{+}_{+1}(t) &= \frac{1}{2} \left(1+ e^{-t \left(2 (\g_{1}+\g_{3})+\g_{10}/2\right)} \times \right. \notag \\
& \ \ \left. \left[ \frac{(8J)^2 e^{-\g_d t/2}+\g_d (2\g_{11}+\g_{12}e^{-\g_d t/2})}{(8J)^2+\g_d^2}\cos(2Jt) \right. \right. \notag \\
&\ \  \left. \left. + \frac{16J \g_{11}(1+e^{-\g_d t/2})}{(8J)^2+\g_d^2}\sin(2Jt)
\right] \right) \\
p^{+}_{++}(t) & = \frac{1}{2} + \frac{1}{(8J)^2+\g_d^2} e^{-t \left(2 (\g_{1}+\g_{3})+\g_{10}/2\right)} \times \notag \\
&\!\!\!\!\!\! \left[ \left( 1+ e^{-t \g_d/2}\right) \left( \frac{1}{4}\cos(2Jt) (8J)^2 + 4\sin(2Jt) J \g_{11} \right)\right. \notag \\
&\!\!\!\!\!\! + \left.  \frac{1}{4}\cos(2Jt)\g_d \left( \g_{12} e^{-t \g_d/2} + 4\g_{11}+\g_{12}\right)
\right] \\
\g_d  &\equiv 2\g_{11}+\g_{12} \ .
\end{align}
\ees
The oscillation frequency is, for all three initial conditions, again equal to $2J$. We illustrate the prediction of this model in Fig.~\ref{fig:5}. For all parameter values we tried, we always find that the ordering of the fidelity oscillation amplitudes is correct for the $s'=0$ and $s'=1$ states, i.e., $s'=0$ has the larger amplitude. However, the $s'=+$ amplitude is always intermediate between the other two, whereas in our experiments this initial state has the lowest amplitude. For this reason we conclude that the present model is also ultimately inadequate, which we attribute to the fact that in reality more than one spectator qubit is directly coupled to the main qubit (see Fig.~\ref{fig:1}). Our full numerical model, which includes all such spectator qubits, does predict the correct ordering.

We note that in our numerical simulations the lowering operator arises naturally as a consequence of the KMS condition, which at the simulated device temperature of $20\;$mK (much smaller than the gap to the first excited state $\ket{1}$ of around $5$ GHz) strongly favors thermal relaxation to the transmon ground state $\ket{0}$.

\section{Detailed analysis of DD in the rotating frame}
\label{app:F}

All the calculations reported in this section are supported by a Mathematica package~\url{https://www.dropbox.com/s/6h6zqjwor8ym76a/dd-calcs.nb?dl=0}, which can be used to reproduce and test all the claims made below.

Let us first write $U(t) = e^{-i\omega_{\rm d}(Z_1+Z_2)t/2}=U_Z(t)\ox U_Z(t)$, where $U_Z(t) = e^{-\frac{1}{2}i\omega_{\rm d}t Z}$. Then the system-only terms of 
\beq
\tilde{H}_{\rm SB}(t) = \sum_{\alpha\beta}g_{\alpha \beta} \left[U(t)\left(\s^{\alpha}\otimes\s^\b\right)U^{\dagger}(t)\right]\otimes B_{\alpha\beta}
\eeq 
can be written as:
\begin{align}
U(t)\left(\s^{\alpha}\otimes\s^\b\right)U^{\dagger}(t) 
 = \s^\a(t) \ox \s^{\b}(t)\ .
\end{align}
Similarly,
\begin{align}
X_2 U(t)\left(\s^{\alpha}\otimes\s^\b\right)U^{\dagger}(t) X_2 = \s^\a(t) \ox \s^{x\b}(t)\ ,
\end{align}
where
\bes
\begin{align}
\s^\a(t) &= U_Z \s^\a U^\dagger_Z(t) \\
\s^{x\b}(t) &= U^\dagger_Z(t) \s^x \s^\b \s^x U_Z(t) \notag\\
&= -(-1)^{\d_{\b x}}U^\dagger_Z(t) \s^\b U_Z(t)\ ,
\label{eq:D3c}
\end{align}
\ees
and where in Eq.~\eqref{eq:D3c} we used 
\bes
\begin{align}
X_2 (I\ox U_Z(t)) X_2 &= I\ox e^{-\frac{1}{2}i\omega_{\rm d}t XZX} \\
& = I\ox e^{\frac{1}{2}i\omega_{\rm d}t Z}  = I\ox U_Z^\dagger(t)\ .
\end{align}
\ees
Explicitly:
\bes
\label{eq:D5}
\begin{align}
\label{eq:D5a}
\s^x(t) &= \left(
\begin{array}{cc}
 0 & e^{-i t \o_{\rm d}} \\
 e^{i t \o_{\rm d}} & 0 \\
\end{array}
\right) = [\s^{xx}(t)]^*\\
\label{eq:D5b}
\s^y(t) &=\left(
\begin{array}{cc}
 0 & -i e^{-i t \o_{\rm d}} \\
 i e^{i t \o_{\rm d}} & 0 \\
\end{array}
\right) = [\s^{xy}(t)]^*\\
\s^z(t) & = -\s^{xz}(t) = \s^z\ .
\label{eq:D5c}
\end{align}
\ees
Recall that after one cycle of pure-X DD the unitary evolution operator is $\tilde{U} (2\tau) =\tilde{U}' (2\tau) + O(\tau^2)$.
Gathering the terms in the exponent of 
\begin{align}
\tilde{U}' (2\tau)&=\exp\left[-i\tau(\omega_{\rm d}-\omega_{q_1})Z_1-i\int_0^{\tau} dt\ \tilde{H}_{\rm SB}(t)\right.\notag\\
&\qquad \left. -i\int_\tau^{2\tau} dt\ X_2\tilde{H}_{\rm SB}(t)X_2\right]\ ,
\end{align}
we thus have:
\bes
\begin{align}
\label{eq:17a}
& -i\int_0^{\tau} dt\ \tilde{H}_{\rm SB}(t)-i\int_\tau^{2\tau} dt\ X_2\tilde{H}_{\rm SB}(t)X_2 \\
&\ \ = -i\sum_{\alpha\beta}g_{\alpha \beta} \int_0^{2\tau} dt\ \s^\a(t) \ox G^\b(t) \otimes B_{\alpha\beta}
\label{eq:17b}
\end{align}
\ees
where 
\begin{subnumcases} {G^\b(t) = }
 \s^{\b}(t) & for $t\in[0,\tau]$\\
\s^{x \b}(t)  & for $t\in[\tau,2\tau]$
\end{subnumcases}
Using Eq.~\eqref{eq:D5a}, the two integrals in Eq.~\eqref{eq:17b} cancel for the $(\a,\b)=(0,z)$ and $(\a,\b)=(z,z)$ cases:
\bes
\begin{align}
& \int_0^{2\tau} dt\ \s^0(t) \ox G^z(t) = \int_0^{\tau} dt\ Z_2 -\int_\tau^{2\tau} dt\ Z_2 = 0 \\
& \int_0^{2\tau} dt\ \s^z(t) \ox G^z(t) = \int_0^{\tau} dt\ ZZ -\int_\tau^{2\tau} dt\ ZZ = 0 \ .
\end{align}
\ees
For $\o_{\rm d}=0$ we recover the standard DD sequence properties in the Schr\"odinger frame, so the pure-X DD sequence also eliminates the $\b=y,z$ cases. 

For $\o_{\rm d}\neq 0$, consider first the $(\a,\b)=(x,z)$ case. In this case we need to calculate 
\beq
I^{xz}\equiv \int_0^\tau dt\ \s^x(t)\ox \s^{z}(t) + \int_\tau^{2\tau} dt\ \s^x(t)\ox \s^{xz}(t)\ .
\eeq 
Changing variables and using Eq.~\eqref{eq:D5c}, we can rewrite this as: 
\beq
I^{xz} = \int_0^\tau dt\ [\s^x(t)\ox \s^z - \s^x(t+\tau)\ox \s^{z}]\ ,
\label{eq:Ixz}
\eeq 
which in general is nonzero. However, using Eq.~\eqref{eq:D5a} we see that this vanishes when $\tau = 2\pi/\o_{\rm d}$. In fact, in this case each term integrates to 0 individually, so the DD sequence is not strictly necessary to cancel $XZ$ at fine-tuned intervals. 

Next, consider the $(\a,\b)=(z,y)$ case.  In this case we need to calculate 
\beq
I^{zy}\equiv \int_0^\tau dt\ \s^z(t)\ox \s^{y}(t) + \int_\tau^{2\tau} dt\ \s^z(t)\ox \s^{xy}(t)\ .
\eeq 
Again changing variables and using Eq.~\eqref{eq:D5c}, we can rewrite this as 
\beq
I^{zy} = \int_0^\tau dt\ \s^z\ox [\s^y(t) + \s^{xy}(t+\tau)] \ .
\label{eq:Izy}
\eeq 
Using Eq.~\eqref{eq:D5b}, we see that the integral vanishes provided $\tau = 2\pi/\o_{\rm d}$; again, the terms also integrate to 0 individually. 

Similarly, we find by explicit calculation (not shown) that:
\begin{itemize}
\item The pairs 
\beq
(\a,\b),(\b,\a)\in\{(0,x),(0,y),(x,z),(y,z)\}
\eeq 
cancel for pulse intervals $\tau$ that are integer multiples of $2\pi/\o_{\rm d}$ both with and without the pure-X DD sequence.
\item The remaining cases 
\beq
(\a,\b)\in\{(0,0),(x,x),(x,y),(y,x),(y,y),(z,0)\}
\eeq 
never perfectly cancel (the $(0,0)$ case is the pure-bath term).
\end{itemize}

In other words, $\omega_{\rm d}$ determines the pulse interval $\tau$ that leads to additional suppression of terms in $\tilde{H}_{\rm SB}(t)$. Namely, the following set of terms also cancels to first order in $\tau$ that is an integer multiple of $2\pi/\o_{\rm d}$: 
\beq
\{X_1,X_2,Y_1,Y_2,XZ,ZX,YZ,ZY\}\ .
\label{eq:suppressed}
\eeq
In general these terms, when integrated, scale as $g_{\a\b} / \o_d$ and so there is also \emph{approximate} cancellation in the limit $\o_d \gg g_{\a\b}$. The remaining terms, $\{Z_1,XX,XY,YX,YY\}$ 
do not perfectly cancel for any value of $\o_{\rm d}$, nor do they have the $g_{\a\b} / \o_d$ scaling (they contain a leading term proportional to $\tau$) and hence are not suppressed even to first order in $\tau$. Unfortunately, we were unable to test the effect of fine-tuning $\tau$ experimentally, since it requires a very precise knowledge of the drive frequency, which appears to be difficult to obtain even with access to OpenPulse~\cite{openpulse}. In any case we are deep in the limit $\o_{\rm d} \gg g_{\a\b}$ ($\omega_{\rm d}\approx 5\;$GHz and $g_{\a\b}\approx 0.1\;$GHz, where we have estimated $\omega_{\rm d}$ via the two-qubit model which gives either $\omega_{\rm d} = \o_q$ or $\omega_{\rm d} = \o_q\pm 2J$; we also have $\o_q \approx 5\;$GHz from Table~\ref{table1} and $J/2\pi\approx 50\;$KHz from fitting our simulations to the experimental data), and so we expect approximate cancellation regardless of $\tau$.

The same reasoning and calculations apply when the DD sequence is replaced by ideal $Y$ pulses instead of $X$, with identical results if $\o_{\rm d}$ is not fine-tuned. When ideal $Z$ pulses are used instead, they do not cancel the $(\a,\b)=(0,z)$ and $(\a,\b)=(z,z)$ cases, since then the pulses commute with the $IZ$ and $ZZ$ interactions. I.e., $Z$ pulses act just like the identity operation on these interactions. 

The XY4 sequence $X f_\tau Y f_\tau X f_\tau Y f_\tau$ is equivalent to $(X f_\tau X) (Z f_\tau Z) (Y f_\tau Y) (I f_\tau I)$. To first order in the pulse interval $\tau$, in the lab frame the XY4 sequence can thus be represented as $\exp[-i\tau (H+XHX+YHY+ZHZ)]$, but in the rotating frame we must break this up into a sum over four integrals, similar to Eq.~\eqref{eq:17a}, and we end up with $\int_0^{\tau} dt [\tilde{H}_{\rm SB}(t) + X_2\tilde{H}_{\rm SB}(t+\tau)X_2+Y_2\tilde{H}_{\rm SB}(t+2\tau)Y_2 + Z_2\tilde{H}_{\rm SB}(t+3\tau)Z_2]$, as mentioned in the main text. Combining all this, we see that when $\o_{\rm d}\neq 0$ is not fine-tuned its effect is identical to that of a pure-X or pure-Y DD sequence, in the sense of exact cancellation. If, on the other hand, we choose pulse intervals $\tau$ that are integer multiples of $2\pi/\o_{\rm d}$ then the pure-X list of suppressed terms [Eq.~\eqref{eq:suppressed}] grows in the XY4 case to include everything except $ZI$ (and $II$, of course). 

Regarding approximate cancellation, now almost all the integrals scale as $g_{\a\b} / \o_d$ (including for $\{XX,XY,YX,YY\}$ where the terms that scale like $\tau$ exactly cancel, unlike in the pure-X case) and so there is also suppression of all terms except $ZI$ and $II$ in the limit $\o_{\rm d} \gg g_{\a\b}$. This is the sense in which the XY4 sequence retains approximate universality.

Since using the XY4 sequence when $\o_{\rm d}\neq 0$ (except fine-tuned values such as $\o_{\rm d}=2\pi/\tau$ or large values $\o_{\rm d} \gg g$) does \emph{not} lead to additional cancellations in the rotating frame of terms that anticommute with $Y$ and $Z$, using the pure-X DD sequence is preferred to XY4 in this case, as the former is shorter by a factor of two. However, as we performed our experiment in the fast drive ($\o_{\rm d} \gg g$) limit, we expect XY4 to perform better when these couplings are present. Indeed, we see pure-X and XY4 sequences perform similarly on most processors but XY4 performs somewhat better on Lima, where relaxation of spectator qubits (likely indicating stronger $\{XX,XY,YX,YY\}$-type couplings to the bath) is significant. This is consistent with the relatively short $T_1$ time (compared to $T_2$) seen for the Lima processor in Table~\ref{table1}; all other processors have $T_1 > T_2$ for all their qubits, with one exception (Q2 of Yorktown on 1/19/2021).

Of course, when $\o_{\rm d}= 0$, i.e., when DD is performed in the lab frame, the XY4 sequence is the usual universal decoupling sequence that cancels all terms involving $X$, $Y$, or $Z$ on the qubit that the pulses are being applied to, while the pure-X sequence only cancels the $Y$ and $Z$ terms. 

Finally, note that applying the DD sequence to the main qubit has the identical effect except for switching the suppressed $Z_2$ system-bath coupling term to $Z_1$. Applying the DD sequence to both qubits does not suppress the crosstalk since $[XX,ZZ]=0$.

\section{Explanation of the non-standard behavior of the pure-X and XY4 sequences in the rotating frame}
\label{app:G}

The analysis of DD sequences is usually carried out in the ``toggling frame'', i.e., the interaction picture defined by the time-dependent Hamiltonian that generates the pulses (see, e.g., Refs.~\cite{Viola:1999:4888,Ng:2011dn}). This is done for mathematical convenience, and we perform such an analysis in Sec.~\ref{app:H} to complement and complete the discussion of DD performance. In this work we chose a different interaction picture motivated by the physics of transmon-based QCs, namely that defined by the uncoupled Hamiltonian $H_{\rm u}=-\frac{1}{2}\sum_i Z_i$, rotating at the frequency $\omega_{\rm d}$, which we identified with the main qubit eigenfrequency. In this frame all operators that do not commute with $H_{\rm u}$ acquire a time dependence, and their rotation frequency is given by $\o_{\rm d}$  [as in Eq.~\eqref{eq:D5}]. Below we explain the reason for this frame choice. But first, let us note that it is this time dependence that prevents the cancelation of terms that would ordinarily cancel under the pure-X or XY4 sequences. For example, both $I^{xz}$ [Eq.~\eqref{eq:Ixz}] and $I^{zy}$ [Eq.~\eqref{eq:Izy}] would ordinarily cancel, since the Pauli operators acting on the spectator qubit in $H_{SB}$ are $Z$ and $Y$, respectively, which anticommute with the $X$-type DD pulses addressed at this qubit. However, in the $I^{xz}$ case, the previously static $X$ operator on the main qubit now has a time dependence, and likewise in the $I^{zy}$ case the previously static $Y$ operator on the spectator qubit now has a time dependence, which appears in $\tilde{H}_{SB}$. This prevents cancellation, except when the time-dependent operators return to the origin, i.e., when $\tau\o_{\rm d}$ is an integer multiple of $2\pi$ (we call this ``fine-tuning $\tau$''). In this case the integral of each of the time-dependent operators vanishes independently, as mentioned in Sec.~\ref{app:F}.
Similarly, when $\omega_{\rm d}$ is large compared to $g_{\a\b}$ and $1/\tau$, the operators rapidly oscillate and so approximately cancel under the rotating wave approximation.

Let us now explain the reason for the particular frame choice we have made.
Consider a single qubit subject to gates or pulses. In the lab frame the Hamiltonian is 
\beq
H_{\rm lab}(t) = -\frac{1}{2}\o_{q}Z + \varepsilon(t) \sin(\omega_{\rm d} t + \phi)X\ , 
\eeq
where $\omega_{\rm d}$ is the drive frequency and $\varepsilon(t)$ is the pulse envelope. If we transform into a frame defined by $U_Z(t) = e^{-\frac{1}{2}i\o_{\rm d}tZ}$, then in this rotating frame the Hamiltonian becomes: 
\bes
\begin{align}
H_{\rm rf}(t) &= -\frac{1}{2}(\o_{q}-\o_{\rm d})Z  \\
&+ \varepsilon(t) \sin(\omega_{\rm d} t + \phi)[\cos(\omega_{\rm d} t)X+\sin(\omega_{\rm d} t)Y] \notag \\
&\approx \frac{1}{2}(\o_{\rm d}-\o_{q})Z +  \frac{1}{2}\varepsilon(t) [\sin(\phi)X+\cos(\phi)Y]\ ,
\label{eq:RWA}
\end{align}
\ees
where in Eq.~\eqref{eq:RWA} we made the rotating wave approximation (discarding terms with a frequency of $\pm 2\o_{\rm d}$).

In an analysis of DD that starts from first principles, such as is common in the nuclear magnetic resonance literature (see, e.g., Ref.~\cite{Alvarez:2010ve}), one assumes that the drive frequency is resonant with the qubit frequency, i.e., $\o_{\rm d} = \o_{q}$. Choosing the phase $\phi$ as $\pi/2$ or $0$ generates an $X$ or $Y$ pulse, respectively, provided one also chooses the pulse width $\d$ appropriately, i.e., such that $\int_0^{\d} \varepsilon(t) dt = \pi$.
One can then realize either a pure-X or an XY4 sequence described by \emph{static} pulses. Thus, in writing an expression such as $X f_\tau Y f_\tau X f_\tau Y f_\tau$ describing the XY4 sequence, one has implicitly assumed a transformation to the frame defined by $U_Z(t)$, along with the resonance condition. 

The situation is slightly more complicated in the presence of a second (spectator) qubit, since the latter can shift the eigenfrequency of the first (main) qubit. This is precisely what happens in the presence of crosstalk, as we saw in the main text. If $\o_{\rm d}$ is not exactly equal to the main qubit's eigenfrequency $\o'_{q}$ then the latter will be subject to a Hamiltonian containing a term of the form $-\frac{1}{2}(\o'_{q}-\o_{\rm d})Z$ in a frame rotating with $\o_{\rm d}$. This mismatch is indeed realized in our experiments, since the gate drive frequency $\o_{\rm d}$ is set once per device after a calibration procedure, while the main qubit's eigenfrequency depends (due to crosstalk) on the state of the spectator qubit. The main qubit is in our case subject to two gates per circuit: a gate that prepares the initial state ($\ket{\pm}$ or $\ket{\pm i}$), and a gate that undoes this preparation before the qubit is measured in the $Z$-basis. The difference between $\o_{\rm d}$ and the main qubit's eigenfrequency then manifests as the oscillations observed in Fig.~\ref{fig:main_1} of the main text, whenever the main qubit's eigenfrequency is shifted by the spectator qubit's initial state so as to be different from $\o_{\rm d}$. 

To sum up, our frame choice is motivated simply by the observation that the drive frequency is calibrated to be resonant with a particular eigenfrequency of the device's qubits, which depends on the state of the neighboring qubits. Transforming into a frame that rotates with one of these frequencies gets us as close as possible to removing the qubit frequency oscillations and hence being able to describe the gates and pulses as static in the rotating frame. 

Next, let us discuss DD in this context. Recall that our setting involves applying DD pulses to the spectator qubits.
The same considerations as above apply. Namely, the drive frequency for the DD pulses is the same $\omega_{\rm d}$ as for the main qubit, since the identification of main \textit{vs} spectator is an arbitrary one we have made. However, the eigenfrequency $\omega'_{q_i}$ ($i>1$) of the spectator qubits depends on the state of the qubits they are coupled to [the main qubit ($i=1$), and possibly other spectator qubits]. This means that generally the resonance condition cannot be satisfied: $\o_{\rm d}\ne \omega'_{q_i}$ for most or all $i$. Therefore, in reality, due to crosstalk we cannot assume that the DD sequence is described by static pulses, and an expression such as $\tilde{U}_X (2\tau) = X_{2} \tilde{U}_f(2\tau,\tau) X_{2} \tilde{U}_f(\tau,0)$ (which is at the core of our analysis of DD in the main text), is an approximation in the sense that the $X_2$ operators should really be replaced by $H_{\rm rf}(t)$ (with $\phi=\pi/2$). The approximation made in our analysis and simulations amounts to assuming, as is common in the analysis of DD, that the pulses are instantaneous, i.e., that the pulse envelope $\varepsilon(t)$ is a Dirac delta function. This is the sense in which $\tilde{U}_X (2\tau)$ can be written as $X_{2} \tilde{U}_f(2\tau,\tau) X_{2} \tilde{U}_f(\tau,0)$, and it is this approximation that is the reason that the spectator qubit frequencies do not appear in our description of DD. With this caveat in mind, the agreement between our simulations and the experimental results shows that the instantaneous pulse approximation works remarkably well.

Now, our analysis in Sec.~\ref{app:F} has already shown that in the relevant rotating frame, even subject to the instantaneous pulse approximation, the effect of the pure-X and XY4 sequences applied to the spectator qubit is equivalent in the sense of exact cancellation (unless $\tau$ is fine-tuned): both cancel the $Z_2$ and $ZZ$ terms, but no others. When $\o_{\rm d} \gg g_{\a\b}$, the XY4 sequence approximately cancels the $XX,XY,YX,YY$ terms that are not suppressed even to first order in $\tau$ by the pure-X sequence. This is also confirmed experimentally in Fig.~\ref{fig:4}, where we see equal performance on a processor where dephasing and $ZZ$ couplings dominate, but improved performance from XY4 on the Lima processor where relaxation is significant. We thus conclude that pure-X DD may be preferable in the circumstance where dephasing and $ZZ$ couplings are dominant, and that both pure-X and XY4 DD should be fine-tuned or operated in the fast-drive limit in order to optimize effectiveness.

\section{Toggling frame analysis}
\label{app:H}

In this section we perform a toggling frame analysis after first transforming to the rotating frame we have considered thus far. DD in the toggling frame is described in terms of sign-switching functions, which lends itself to a simpler interpretation. As our open system model we consider a single qubit coupled to an external bath described by the following Hamiltonian: 
\bes
\begin{align}
    H &= -\frac{1}{2}\omega_qZ + H_{\mathrm{DD}}(t)+ H_\mathrm{B} \\
    &\quad+ g_xXB_X + g_yYB_Y + g_zZB_Z + H_\mathrm{B} \ ,
\end{align}
\ees
where $H_\mathrm{DD}(t)$ is the Hamiltonian generating the DD control pulses. We choose 
\bes
\begin{align}
    H_\mathrm{DD}(t) &= \epsilon_X(t)(\cos(\omega_{\rm d} t)X-\sin(\omega_{\rm d} t)Y) \\
    &\quad+\epsilon_Y(t)(\sin(\omega_{\rm d} t)X+\cos(\omega_{\rm d} t)Y) \ ,
\end{align}
\ees
where $\omega_{\rm d}$ is the drive frequency and $\epsilon_X(t)$ and $\epsilon_Y(t)$ are the pulse envelopes. 

If we move to a rotating frame generated by
$-\frac{1}{2}\omega_{\rm d} Z + H_\mathrm{B}$ (thus also including the pure-bath term in addition to our standard rotating frame),
then the interaction picture Hamiltonian becomes:
\bes
\begin{align}
    &\tilde{H}(t) = - \frac{1}{2}(\omega_q-\omega_{\rm d})Z+ \epsilon_X(t)X + \epsilon_Y(t)Y\\
    &\quad+ g_xX(t)B_X(t)+g_yY(t)B_Y(t) + g_zZB_Z(t) \ ,
\end{align}
\ees
where
\bes
\begin{align}
    X(t) &=\cos(\omega_{\rm d} t) X + \sin(\omega_{\rm d} t) Y \\
    Y(t) &=\cos(\omega_{\rm d} t) Y - \sin(\omega_{\rm d} t) X \\
    B_i(t) & = U^\dagger(t) B_i U(t) \ ,
\end{align}
\ees
where $U(t) = e^{-it(-\frac{1}{2}\omega_{\rm d} Z + H_\mathrm{B})}$.
If the pulses are ideal, we can approximate $\epsilon_X(t)$ and $\epsilon_Y(t)$ by Dirac comb functions:
\bes
\begin{align}
    \epsilon_X(t) &= \frac{\pi}{2}\sum_{i=0}^{N_X-1}\delta(t-t^x_i) \\
    \epsilon_Y(t) &= \frac{\pi}{2}\sum_{i=0}^{N_Y-1}\delta(t-t^y_i) \ ,
\end{align}
\ees
where $t^x_i$ and $t^y_i$ are the locations of the $X$ and $Y$ pulses and $N_X$ and $N_Y$ are the numbers of each pulse type. Then we can move to the toggling frame~\cite{Viola:1999:4888} defined by the pulse unitaries
\bes
\begin{align}
    U_{R}(t)\propto\left\{\begin{aligned}
    R, \quad & t_{2 i}^{r} \leq t<t_{2 i+1}^{r} \\
    I, \quad & t_{2 i+1}^{r} \leq t<t_{2 i+2}^{r} \\
    I, \quad & t<t_0
    \end{aligned}\right.
\end{align}
\ees
up to a global phase. In the above expression, $R\in[X, Y]$ and accordingly $r\in [x, y]$. If we assume $t^x_i \neq t^y_j$ for all $i, j$, then $[U_X(t), U_Y(t)] = 0$. The toggling frame is defined by the unitary $U_X(t)U_Y(t)$.

The Hamiltonian in the toggling frame becomes:
\begin{align}
    \tilde{H}_{\rm tog}(t) &= f_X(t)f_Y(t)\big[- \frac{1}{2}(\omega_q-\omega_{\rm d})Z + g_z ZB_Z(t)\big] \notag \\
    & \quad + g_xf_Y(t)X(t)B_X(t)+g_yf_X(t)Y(t)B_Y(t) \ ,
\end{align}
where $f_X(t)$ and $f_Y(t)$ are switching functions given by:
\beq
    f_{R}(t)=\left\{\begin{aligned}
-1, & \quad t_{2 i}^{r} \leq t<t_{2 i+1}^{r} \\
1, & \quad t_{2 i+1}^{r} \leq t<t_{2 i+2}^{r} \\
1, & \quad t<t_0
\end{aligned}\right. \ .
\eeq
Next, we focus only on equidistant DD sequences. We denote the pulse interval by $\tau$ and the length of a single DD cycle by $\Delta t \propto \tau$. Then we calculate the total unitary evolution operator in the toggling frame from $0$ to $\Delta t$ using the first order Magnus expansion:
\bes
\begin{align}
    &U_{\rm tog}(\Delta t) \approx \exp\bigg[-i\int_0^{\Delta t} dt \tilde{H}_{\rm tog}(t) \bigg] = \\
    & \exp\bigg[-i\int_0^{\Delta t} dt \bigg(\frac{f_X(t)f_Y(t)}{2}(\omega_{\rm d}-\omega_q)Z\label{eq:closed_system_part}\\
    &\quad +f_X(t)f_Y(t)g_z ZB_Z(t)+g_xf_Y(t)X(t)B_X(t)\\
    &\quad +g_yf_X(t)Y(t)B_Y(t) \bigg) \bigg] \ .
\end{align}
\ees
We ignore the terms in line~\eqref{eq:closed_system_part} because they describe the closed-system dynamics. The remaining terms in the exponential can be written as:
\begin{equation}
    \label{eq:DD_error}
    \int_0^{\Delta t} dt [G_X(t)X + G_Y(t)Y + G_Z(t)Z] \ ,
\end{equation}
where
\bes
\begin{align}
    G_X(t) & = g_xf_Y(t)\cos(\omega_{\rm d} t) B_X(t) - g_yf_X(t)\sin(\omega_{\rm d} t) B_Y(t) \label{eq:GX}\\
    G_Y(t) & = g_xf_Y(t)\sin(\omega_{\rm d} t) B_X(t) + g_yf_X(t)\cos(\omega_{\rm d} t) B_Y(t) \label{eq:GY}\\
    G_Z(t) &= g_zf_X(t)f_Y(t)B_Z(t) \label{eq:GZ} \ .
\end{align}
\ees
To achieve first order cancellation, we need to have
\begin{equation}
    \label{eq:DD_target}
    \int_0^{\Delta t} G_R(t) dt = 0 \quad R\in[X,Y,Z] \ .
\end{equation}
This is not always possible when $\omega_{\rm d} \neq 0$, even for a universal DD sequence such as XY4. A counterexample is given in Sec.~\ref{app:F}. Recall that in Sec.~\ref{app:G} we argued that in the limit of $\omega_{\rm d} \gg g$, Eq.~\eqref{eq:DD_target} is still approximately achievable.

We now derive an upper bound for Eq.~\eqref{eq:DD_error} under two different assumptions. First, we consider the case where the bath is constant, i.e., $B_R(t)=B_R$. In this case it is easy to see that $\int_0^{\Delta t} G_Z(t) dt$ (which has no drive frequency dependence) can be exactly cancelled. We bound the remaining terms by first using the triangle inequality:
\begin{align}
    &\left\lVert\sum_R R\int_0^{\Delta t} G_R(t) dt\right\rVert \notag \\
    &\quad\leq \lvert g_x \rvert \left\lVert B_X \right\rVert \left\lvert\int_0^{\Delta t} f_Y(t)\cos(\omega_{\rm d} t) dt \right\rvert + \cdots \ .
    \label{eq:sum_error}
\end{align}
Note that in the absence of the $\cos(\omega_{\rm d} t)$ term the integral would vanish for any choice of $\Delta t$ that includes an equal number of $\pm 1$ switches, as in standard DD. It is thus the presence of the drive frequency that causes the imperfect cancellation, as we have argued above. On the other hand, it is also clear that a sufficiently large drive frequency can average the integral to zero. To show this rigorously, terms such as $\left\lvert\int_0^{\Delta t} f_Y(t)\cos(\omega_{\rm d} t) dt \right\rvert$ can be bounded by splitting the integral into intervals over which the switching functions are constant:
\begin{align}
    \left\lvert\int_0^{\Delta t} f_Y(t)\cos(\omega_{\rm d} t) dt \right\rvert \leq \sum_i\left\lvert\int_{\tau_i}^{\tau_{i+1}}\cos(\omega_{\rm d} t) dt \right\rvert \leq \frac{2c'}{\omega_{\rm d}} \ ,
\end{align}
where $c'$ is the number of piece-wise constant intervals of the function $f_Y(t)$ in $[0, \Delta t]$. In the case of pure-X, pure-Y, or XY4 sequences, it equals $2$. Denoting
\begin{equation}
    g = \max_{r\in\{x,y\}}{\lvert g_r \rvert}, \quad B = \max_{R\in\{X,Y\}}{ \left\lVert B_R \right\rVert} \ ,
\end{equation}
we can add up all terms in Eq.~\eqref{eq:sum_error}:
\begin{equation}
    \left\lVert\sum_R R\int_0^{\Delta t} G_R(t) dt\right\rVert\leq \frac{16gB}{\omega_{\rm d}} \ .
\end{equation}
We can see that the upper bound of the oscillating terms scales as $O(gB/\omega_{\rm d})$. This can also be understood using the Riemann-Lebesgue lemma. Thus, when $\omega_{\rm d} \gg g$, the fast drive already offers some protection against the $X$ and $Y$ coupling to the bath. The system bath coupling along the $Z$ direction becomes the dominant one, which can be suppressed via pure-X, pure-Y, or XY4 sequences.

Second, let us consider the case where the bath is rotating at the same frequency as the drive. In this case, $B_R(t)$ can be written as a Fourier series with only a single frequency component at $\omega_{\rm d}$:
\begin{equation}
    B_R(t) = B'_R \cos(\omega_{\rm d} t) + B''_R\sin(\omega_{\rm d} t) . 
\end{equation}
Based on the same argument, the terms left in Eq.~\eqref{eq:DD_error} that do not depend on $\o_{\rm d}$ are
\bes
\label{eq:H18}
\begin{align}
    \frac{g_xXB'_X}{2}\int_0^{\Delta t} f_Y(t) dt, & -\frac{g_yXB''_Y}{2}\int_0^{\Delta t} f_X(t) dt \\
    \frac{g_xYB''_X}{2}\int_0^{\Delta t} f_Y(t) dt, & -\frac{g_yYB'_Y}{2}\int_0^{\Delta t} f_X(t) dt \ ,
\end{align}
\ees
while the other terms scale as $O(gB/\omega_{\rm d})$. As a result, the fast drive will suppress all the error terms to order $O(gB/\omega_{\rm d})$ except for $XB_X'$, $XB''_Y$, $YB''_X$ and $YB'_Y$. However, the latter cancel (to first order in $\Delta t$) by either $X$ or $Y$ pulses, since the integrals in Eq.~\eqref{eq:H18} all vanish for an appropriate choice of $\Delta t$. 

When the bath has different frequency components the analysis become more complex. However, we expect the results to hold for baths that are concentrated around $\omega = 0$ or $\omega = \omega_{\rm d}$. On the other hand, this analysis shows that the rotating frame introduces an additional non-vanishing order to any DD sequence. For universal DD sequences such as XY4 with leading cancelling order $O(\Delta t)$, some error of the order $O(gB/\omega_{\rm d})$ still remains. This may dramatically affect the performance of DD sequences which are designed with more than first order cancellation (such as CDD~\cite{Khodjasteh:2005xu} or UDD~\cite{Uhrig:2007qf}), because the $O(gB/\omega_{\rm d})$ terms will become comparable to high order errors $O(\Delta t^n)$ when $n$ is large enough.


%

\end{document}